\documentclass[twocolumn,showpacs,preprintnumbers,amsmath,amssymb,APSl,prd,nofootinbib,superscriptaddress]{revtex4-1}

\usepackage{bm}
\usepackage{mathrsfs}
\usepackage{xcolor,color,graphicx,graphics}
\usepackage[all]{xy}
\usepackage{epsfig,subfigure}
\usepackage{latexsym,amssymb,amsmath,amsfonts} 
\usepackage[english]{babel} 
\usepackage[OT1]{fontenc}
\usepackage[latin1]{inputenc}
\usepackage{makeidx}
\usepackage{hyperref}
\usepackage{color,graphicx,graphics,wrapfig,epsf}
\usepackage{calligra}

\newcommand{\be}{\begin{equation}}
\newcommand{\en}{\end{equation}}
\newcommand{\bea}{\begin{eqnarray}}
\newcommand{\ena}{\end{eqnarray}}

\begin{document}

\title{Structure and thermodynamics of charged non-rotating black holes in higher dimensions}

\author{H. Benbellout}\email{astrohamza@gmail.com}
\affiliation{LPTM, Universit\'e de Cergy-Pontoise. 2 Av. Adolphe Chauvin. 95302 Cergy Pontoise, France.}
\author{J. Diaz-Alonso} \email{joaquin.diaz@obspm.fr}
\affiliation{LUTH, Observatoire de Paris, PSL Research University, CNRS, Universit\'e Paris
Diderot, Sorbone Paris Cit\'e. 5 Place Jules Janssen, 92190 Meudon, France.}
\affiliation{Departamento de F\'isica, Universidad de Oviedo. Avenida
Calvo Sotelo 18, 33007 Oviedo, Asturias, Spain.}
\author{D. Rubiera-Garcia} \email{drgarcia@fc.ul.pt}
\affiliation{Instituto de Astrof\'isica e Ci\^encias do Espa\c{c}o, Universidade de Lisboa,
Faculdade de Ci\^encias,
Campo Grande, PT1749-016 Lisboa, Portugal}

\date{\today}

\begin{abstract}

We analyze the structural and thermodynamic properties of $D$-dimensional ($D \geq 4$), asymptotically flat or Anti-de-Sitter, electrically charged black hole solutions, resulting from the minimal coupling of general nonlinear electrodynamics to General Relativity. This analysis deals with static spherically symmetric (elementary) configurations with spherical horizons. Our methods are based on the study of the behaviour (in vacuum and on the boundary of their domain of definition) of the Lagrangian density functions characterizing the nonlinear electrodynamic models in flat spacetime. These functions are constrained by some admissibility conditions endorsing the physical consistency of the corresponding theories, which are classified in several families, some of them supporting elementary solutions in flat space which are non topological solitons. This classification induces a similar one for the elementary black hole solutions of the associated gravitating nonlinear electrodynamics, whose geometrical structures are thoroughly explored. A consistent thermodynamic analysis can be developed for the subclass of families whose associated black hole solutions behave asymptotically as the Schwarzschild metric (in absence of a cosmological term). In these cases we obtain the behaviour of the main thermodynamic functions, as well as important finite relations among them. In particular, we find the general equation determining the set of extreme black holes for every model, and a general Smarr formula, valid for the set of elementary black hole solutions of such models. We also consider the one-parameter group of scale transformations, which are symmetries of the field equations of any nonlinear electrodynamics in flat spacetime. These symmetries are respected by the minimal coupling to gravitation and induce representations of the group in the spaces of solutions of the different models, characterized by their thermodynamic functions. Exploiting this fact we find the expression of the equation of state of the set of black hole solutions associated to any model. These results are generalized to asymptotically Anti-de-Sitter solutions.

\end{abstract}

\maketitle

\section{Introduction} \label{sectionI}

For decades the study of the structural and thermodynamic properties of black hole (BH) configurations obtained from the coupling of nonlinear electrodynamic (NED) models to the gravitational field in $D$($\geq4$) spacetime dimensions (without or with a cosmological term, leading to asymptotically flat or Anti-de-Sitter (AdS) configurations, respectively) has become a useful tool in the investigation of some fundamental issues, such as the AdS/CFT correspondence \cite{adscft1,adscft2,adscft3}, the quest for regular solutions (see \cite{Ansoldi:2008jw} and references therein), or the investigation of first-order phase transitions in BH thermodynamics \cite{chamblin99,chamblin99b,VdW1,VdW2,VdW3}.

The interest on NED models was originally triggered by the introduction in 1934 of the Born-Infeld model \cite{BI34} as a nonlinear generalization of $D = 4$ Maxwell electrodynamics. The now familiar square-root structure of the Lagrangian density of this model sets a bound on the electric field by which the problem of the divergent self-energy of the classical field of point-like charges is removed. When coupled to gravity, this model yields new geometrical and thermodynamical properties for the corresponding BH solutions, as compared to those of the Reissner-Nordstr\"om solution of the Einstein-Maxwell field equations \cite{BIGR1,BIGR2,BIGR3}. Besides its remarkable properties such as electric-magnetic duality \cite{BIGR2} or its exceptional behaviour regarding wave propagation and absence of birefringence phenomena \cite{WP,WP2}, the interest in this model is also due to the fact that (Abelian and non Abelian) Born-Infeld-like actions, coupled to gravity, naturally arise in the low-energy regime of string theory and D-Brane physics \cite{ST1,ST2,ST3,ST4}. A second meaningful example (in $D = 4$) is the Euler-Heisenberg model \cite{EH36,EH36b}, which arises as an effective Lagrangian of Quantum Electrodynamics, introducing nonlinear corrections to the Maxwell action  which describe, at a classical level, quantum vacuum polarization effects on the photon propagation at leading order in the perturbative expansion \cite{EHQED1, EHQED2}. When minimally coupled to gravity, this model gives a first approach to the corrections introduced by these vacuum effects on the structure of the Reissner-Nordstr\"om BHs generated by bare point-like charges, which could provide observational signatures in  astrophysical contexts \cite{EHGR1,EHGR2,EHGR3}.

These two models are just the tip of a much larger phenomenology regarding the study of NEDs in the gravitational context, which has extended so far to modifications on the geometric \cite{NEDstructure1,NEDstructure2,NEDstructure3,NEDstructure4,NEDstructure5,NEDstructure6,NEDstructure7,NEDstructure8} and thermodynamic properties \cite{NEDthermodynamics1,NEDthermodynamics2,NEDthermodynamics3,NEDthermodynamics4,NEDthermodynamics5} of BHs, generalizations to higher dimensions and to asymptotically cosmological spacetimes \cite{NEDads1,NEDads2,NEDads3,Herdeiro:2015vaa,Peca:1998cs,Hendihig,Dehyadegari:2017hvd,Kuang:2018goo}, search of models with regular elementary BH solutions \cite{NEDregular1,NEDregular2,NEDregular3,NEDregular4,NEDregular5,NEDregular6,NEDregular7,NEDregular8,Rodrigues:2018bdc}, wave propagation in these backgrounds \cite{NEDpropagation,Perlick:2018rup}, or light-by-light scattering phenomena \cite{Ellis:2017edi,Rebhan:2017zdx}, among many others. Some of these models and their associated solutions have been further discussed within the context of gravitational extensions of General Relativity \cite{BeyondGR1,BeyondGR2,BeyondGR3,BeyondGR4,BeyondGR5,BeyondGR6,BeyondGR7}. However, most of the available literature so far has focused on particular NED models, selected either on fundamental grounds or as phenomenological tools to address diverse theoretical, astrophysical and cosmological problems, while general analysis of these scenarios are still scarce.

In a couple of previous works \cite{dr10a,dr10b} two of us introduced general methods for the systematic and exhaustive analysis of the geometrical structures of the elementary BH solutions associated to general NEDs minimally coupled to gravity in $D = 4$ spacetime dimensions. In a flat spacetime these models are characterized by Lagrangian densities which are arbitrary functions $\varphi(X,Y)$ of the two quadratic field invariants, $X \equiv -\frac{1}{2} F_{\mu\nu}F^{\mu\nu}$, $Y \equiv -\frac{1}{2} F_{\mu\nu}F^{*\mu\nu}$, that can be built out of the field strength tensor $F_{\mu\nu} = \partial_{\mu}A_{\nu}-\partial_{\nu}A_{\mu}$ and its dual $F^{*\mu\nu} = \frac{1}{2}\epsilon^{\mu\nu\alpha\beta}F_{\alpha\beta}$, where $A_{\mu}$ is the four-vector potential. These models were constrained by several conditions in order to obtain physically consistent theories. Such conditions include regularity of the Lagrangian function $\varphi(X,Y)$ on its open and connected domain of definition, positivity of the energy, and parity invariance. With these constraints, the heart of such methods lies on a classification of the NED models into several families, which are characterized by the central and asymptotic behaviours of their elementary solutions (or, equivalently, by the behaviour of the Lagrangian densities in vacuum and near the boundary of their domain of definition around $Y = 0$ in the $X-Y$ plane, regardless of their explicit forms elsewhere in this domain). This way, once such behaviours are known, one can fully characterize the geometric structure of the BH solutions corresponding to a given family from a qualitative point of view, while the specification of the full expression of the particular Lagrangian density in the family allows to establish the quantitative details. Using these methods we also found a number of novel results in the general thermodynamic analysis of these models \cite{dr13}. Among them we underline the finding of a generalized version of the Smarr formula \cite{smarr73} holding for any gravitating NED (G-NED) (containing the several Smarr formulae obtained in the literature for particular cases), and the investigation of some consequences on BH thermodynamics of the scale invariance laws of NEDs in flat space, which are respected for the elementary charged BH solutions when minimally coupled to gravity, introducing large simplifications in the analysis of this issue.

The aim of the present paper is to carry out a detailed extension of the above methods and results to BH configurations supported by NEDs in $D \geq 4$ spacetime dimensions minimally coupled to gravity, in both asymptotically flat and asymptotically AdS backgrounds. The consideration of this extension is of interest from the point of view of the applications of NEDs within the context of the AdS/CFT correspondence (see e.g. \cite{BI-holo1,BI-holo2,BI-holo3,BI-holo4,BI-holo5,BI-holo6}). The analysis carried out here will be able to collect, classify and describe into a single framework most of the examples considered in the literature so far. This way we will be able to characterize the new geometric and thermodynamic features of the corresponding BHs, and to compare them to those obtained in the asymptotically flat $D = 4$ cases. Besides the contribution to the improvement of the understanding of the geometric and thermodynamic properties of BHs in $D\geq4$, one of the main novelties of the results presented here is their broad generality, since only a few constrains are imposed upon the Lagrangian densities in order to deal with physically consistent theories on the matter sector.

This work is organized as follows:

In section \ref{sectionII} we define the families of models considered as well as our conventions. After introducing the admissibility constraints, we classify the models, in $D \geq 4$ flat spacetime, in terms of the behaviour of their Lagrangian densities in vacuum and around the boundary of their domain of definition which, as in the $D=4$ cases, are shown to correspond to the asymptotic and central field behaviours of their elementary solutions, respectively. The results of this section generalize the four-dimensional analysis of Refs. \cite{dr10a,dr10b}.

Section \ref{sectionIII} is devoted to the study of the elementary solutions of the Einstein equations resulting from the minimal coupling of generalized admissible NEDs to gravitation in $D \geq 4$ (with and without a cosmological term), restricted here to those exhibiting topologically spherical horizons. We analyze the corresponding geometric structures of these solutions, both in asymptotically flat and AdS backgrounds, with special emphasis on the characterization of the horizons for the different families. We consider also the set of extreme BHs associated to a given model and obtain a general formula fully characterizing it.

In section \ref{sectionIV} we carry out the thermodynamic study of the asymptotically flat and AdS black hole solutions. In the asymptotically flat cases, and for those families for which a first law of BH thermodynamics can be consistently introduced, we define the main thermodynamic functions and obtain the qualitative form of the phase diagrams, which give the horizon structures for the BH solutions of the different families in the charge-mass plane. Moreover, we obtain the behaviour of these thermodynamic functions under the action of the scale transformations, generalizing the results obtained $D = 4$ spacetime dimensions. Next, these results are extended to the case of asymptotically AdS black holes. Whenever an ambiguity arises concerning the asymptotically flat or AdS character of some thermodynamic variables (mainly the mass and the temperature), they will be characterized via a subindex $AF$ or $AdS$.

Section \ref{sectionV} deals with the analysis of the relations between the thermodynamic functions. We will obtain a generalized expression of the Smarr law \cite{smarr73}, valid for all the G-NEDs in any $D\geq4$ dimensions, which reduces to the  expressions found in the literature for a few particular cases. Next, this law is further generalized to asymptotically AdS black holes. Special attention is paid to the group structure underlying the scale invariance of NED models. The representations of this group in the spaces of BH solutions, characterized by their thermodynamic functions, allow to obtain universal relations between such functions and their derivatives, which correspond to the generating equations of the group representations in the different (three-dimensional) spaces of state variables (in fact, the generalized Smarr formula is shown to be equivalent to the generating equation of the group representation in the charge-entropy-mass space). The beams of characteristics of these equations define the group trajectories, which are independent of the particular models. These characteristics generate the sets of BH solutions of the different models as two-dimensional surfaces in those three-spaces. In the particular case of the charge-entropy-temperature space, the extreme BH equations allow the explicit determination of the \emph{equation of state} (EOS) for the full set of BH solutions associated to any model. The corresponding two-dimensional surfaces in this space contain the full thermodynamic information on the ensembles of BH solutions of different models.

It should be stressed that our thermodynamic analysis concerns the ensembles of BH-states which are the sets of elementary solutions of the different particular G-NEDs, characterized by the usual state variables (mass, charge, temperature, entropy, etc.). It excludes the extensions for which the cosmological constant \cite{Teitelboim1985,kastor2009} or some internal parameters of the NED Lagrangian densities \cite{mann2012} are treated also as state variables. Nevertheless, some aspects of these extensions for which our results on the scale behaviours are pertinent, will be discussed at the end of section \ref{sectionV}.

We conclude in section \ref{sectionVI} with a discussion and some perspectives for future research.

\section{General nonlinear electrodynamics in flat $D \geq 4$ spacetime dimensions} \label{sectionII}

This section will establish the basic framework upon which the subsequent analysis of G-NEDs will be carried out. Therefore, we shall develop it with some detail. Let us then consider NEDs in flat $D \geq 4$ spacetime dimensions, whose dynamics is governed by Lagrangian densities defined as functions of the unique \textit{quadratic} invariant which can be built from the field strength tensor in all these cases as:
\be
\mathcal{L} = \varphi(X) \hspace{0.1cm}; \hspace{0.1cm} X = -\frac{1}{2} F_{\mu \nu}F^{\mu \nu} \ .
\label{eq:(2-1)}
\en
Hereafter, Greek indices run from $0$ to $d = D - 1$ and Latin indices run from $1$ to $d = D-1$. We exclude in the Lagrangian density dependencies on more complex objects which can be built from the tensor field. The invariant $X$ can be explicitly written as
\be
X = \sum_{i} (F_{0i})^{2} - \sum_{i>j} (F_{ij})^{2} = (\vec{E})^{2} - \sum_{i>j} (F_{ij})^{2} \ .
\label{eq:(2-1)bis}
\en
This defines the electric field as a $(D-1)$-vector whose components are $E_{i}\equiv F_{0i}$. The ``magnetic" components defined from $F_{ij}$ have now a tensorial character in $D-1$ space dimensions.

A number of constrains are now introduced on the Lagrangian density functions. First, we require them to be defined in an open and connected domain of the $X$-axis, including the ``vacuum" ($X = 0$). Second, we require $\varphi(X)$ to be at least of class $C^{1}$ on its domain of definition, with the possible exception of $X = 0$, where it is assumed to be at least of class $C^{0}$. Finally, we shall require the positivity of the energy density for any field. The explicit form of the latter constraint will be specified in section \ref{sec:IIB}. These requirements are regarded as minimal conditions for physical consistency of the corresponding theories, defining what we shall call hereafter \emph{admissible} NED models (see Ref.\cite{dr09} for a more detailed discussion on admissibility conditions).

The full action for the electromagnetic field including the currents is given by
\be
\mathcal{S} = \int d^D x \left[ \varphi (X) - \xi A_{\mu} J^{\mu} \right]\ ,
\label{eq:(2-1)ter}
\en
where the constant $\xi$ allows to fix the units of charge.

\subsection{The field equations}

The field equations resulting from the action (\ref{eq:(2-1)ter}) for free fields are
\be
\partial_{\mu} [\varphi_X F^{\mu \nu}] = 0 \ ,
\label{eq:(2-2)}
\en
where $\varphi_X \equiv \frac{d \varphi}{d X}$. In presence of external currents these equations pick up a new term of the form
\be
\partial_{\mu} [\varphi_X F^{\mu \nu}] = \frac{\xi}{2} J^{\nu} \ ,
\label{eq:(2-3)}
\en
where $J^{\nu}$ is the current $D$-vector. The total charge of a given distribution is defined as
\be
Q = \int d^{D-1}\vec{x} J^{0}(x^{\mu}) \ ,
\label{eq:(2-4)}
\en
and in the static spherically symmetric cases ($J^{0}(x^{\mu}) = J^{0}(r)$, with $r^{2} = \sum_{i=1}^{D-1}(x^{i})^{2}$) this integral takes the form
\be
Q = \omega_{(D-2)} \int_{0}^{\infty} dR R^{D-2}J^{0}(R) \ ,
\label{eq:(2-5)}
\en
where
\be
\omega_{(D-2)} = \frac{2\pi^{(D-1)/2}}{\Gamma[(D-1)/2]} \ ,
\label{eq:(2-5)+}
\en
is the measure of a unit $S^{D-2}$ hypersphere. Let us consider now the case of point-like charges of magnitude $Q$ at rest at the origin, as sources of the field \cite{schw}. In this case the charge density is given by a Dirac-delta distribution, $J^{0} = Q \delta_{D-1}(\vec{r})$. By integrating both sides of Eq.(\ref{eq:(2-3)}) inside the hypersphere $S^{D-2}(r)$ in this electrostatic spherically symmetric (ESS) case we obtain
\be
\omega_{(D-2)} r^{D-2} \varphi_X E(r) = \frac{\xi}{2} Q \ .
\label{eq:(2-5)bis}
\en
With the choice $\xi = 2\omega_{(D-2)}$, which fixes the charge units for each dimension, we obtain a first integral of the field equations as
\be
r^{D-2} \varphi_X E(r) = Q \ .
\label{eq:(2-5)quart}
\en
Eq.(\ref{eq:(2-5)quart}) allows us to obtain the central field $E(r,Q)$ once the explicit form of the Lagrangian density $\varphi(X \equiv E^{2})$ is specified. This expression is the generalization to the $D-$dimensional case of the first integral obtained in $D=4$ \cite{dr09}. The form of the ESS field in terms of the vector potential in the Lorentz gauge ($\vec{A} = 0; A_{0} = A_{0}(r)$) is
\be
\vec{E}(\vec{r}) = E(r) \frac{\vec{r}}{r} = - \vec{\nabla}A_{0}(r) = -\frac{d A_{0}(r)}{dr} \frac{\vec{r}}{r} \ .
\label{eq:(2-5)penta}
\en

\subsection{The energy-momentum tensor} \label{sec:IIB}

The mixed components of the symmetric energy-momentum tensor which result from the Lagrangian density (\ref{eq:(2-1)}) read
\be
{T_\mu}^{\nu} = 2 \varphi_X F_{\mu \beta} F^{\beta \nu} - \varphi \delta_{\mu}^{\nu} \ ,
\label{eq:(2-6)bis}
\en
and its trace takes the form
\be
{T_\mu}^{\mu} = 4 \varphi_XX - D \varphi \ ,
\label{eq:(2-6)ter}
\en
which, in the case of $D-$dimensional Maxwell theory, becomes ${T_\mu}^{\mu} = (4 - D) X$. Thus,  the traceless character of the energy-momentum tensor for the Maxwell theory is only fulfilled in $D = 4$. The general family of models with traceless energy-momentum tensors can be easily obtained from Eq.(\ref{eq:(2-6)ter}). The form of their Lagrangian densities are rational powers of the invariant $X$ and read
\be
\varphi(X) \propto X^{D/4} \ .
\label{eq:(2-6)quinc}
\en
Some of these power-field Lagrangian models coupled to gravity and their ESS solutions have been studied in the literature (see e.g. \cite{Hassaineetal}).

We can now determine the conditions to be satisfied by the Lagrangian density, $\varphi(X)$, in order to implement the requirement of positivity of the energy density. From (\ref{eq:(2-6)bis}), the energy density ${T_0}^{0}$ takes the form
\be
\rho =  {T_0}^{0} = 2 \varphi_X F_{0 \beta} {F^\beta}_{0} - \varphi =
2 \varphi_X \vec{E}^{2} - \varphi \ ,
\label{eq:(2-7)}
\en
where we have used the definition of the electric field in terms of the components of the tensor field ($E_{i} \equiv F_{0 i}$). We require first the energy density to reach its minimum value in vacuum (where $\vec{E} = 0$ and $X = 0$). This minimum can be taken to be zero, without loss of generality, and thus this requirement leads to
\be
\varphi(0) = 0 \ ,
\label{eq:(2-8)}
\en
as a necessary condition. Because the norm of the electric field may take arbitrary large values, another necessary condition is
\be
\varphi_X > 0, \hspace{.3cm} (\forall X \neq 0) \ ,
\label{eq:(2-9)}
\en
which means that $\varphi$ is a strictly monotonically increasing function (excepting in vacuum, where its derivative may vanish). Moreover, if we consider field configurations for which $X < 0$, it is obvious from Eq.(\ref{eq:(2-7)}) that the positivity of the energy requires
\be
\varphi(X < 0) < 0 \ .
\label{eq:(2-10)}
\en
For field configurations with $X > 0$ we have instead the condition
\be
\rho = 2 \varphi_X \vec{E}^{2} - \varphi \geq 2 \varphi_X X - \varphi \geq 0 \ .
\label{eq:(2-11)}
\en
This implies that the function
\be
\frac{\varphi(X)}{\sqrt{X}} \ ,
\label{eq:(2-12)}
\en
must be a positive increasing one for any $X > 0$.

We conclude that the conditions (\ref{eq:(2-9)}), (\ref{eq:(2-10)}) and (\ref{eq:(2-11)}) are necessary and sufficient for the positivity of the energy, and must be satisfied by the Lagrangian density of any admissible model.

\subsection{The elementary solutions and their classification}

Let us come back to the first integral (\ref{eq:(2-5)quart}), where $Q$ is a integration constant identified as the electric charge. The form of this first integral shows that the field depends on its arguments trough the ratio $r/Q^{\frac{1}{D-2}}$. As a consequence, the electrostatic field scales as
\be
E(r,Q,D) = E\left( \frac{r}{Q^{\frac{1}{D-2}}},1,D\right) \ ,
\label{eq:(2-13)}
\en
or, equivalently, as
\be
E(r,Q,D) = E(\theta r,\theta^{D-2}Q,D) \ ,
\label{eq:(2-14)}
\en
$\theta$ being an arbitrary positive parameter. In fact, this is a consequence of the well known invariance of the field equations (\ref{eq:(2-3)}) under the scale transformations
\be
x^{\mu} \rightarrow \theta x^{\mu}\hspace{0.1cm};\hspace{0.1cm} A^{\mu} \rightarrow \theta A^{\mu}\hspace{0.1cm};\hspace{0.1cm} J^{\mu} \rightarrow \theta^{-1} J^{\mu} \ .
\label{eq:(2-15)}
\en
Denoting as $\Gamma(\theta)$ ($\theta>0$) the elements of the one-parameter set of these transformations, it is obvious that it exhibits a one-parameter multiplicative group structure with respect to the product law ($\circ$) of iteration of the transformations:
\be
\Gamma(\theta_{1}) \circ \Gamma(\theta_{2}) = \Gamma(\theta_{1} \cdot \theta_{2}) \hspace{0.1cm};\hspace{0.1cm}  \Gamma(\theta=1) = I \ ,
\label{eq:(2-15)V}
\en
$I$ being the identity transformation. As we shall see in Section \ref{sectionV}, the representations of this scale group will be at the root of useful scale symmetries of the thermodynamic state functions of the elementary G-NEDs black holes.

Let us now establish a classification of the NEDs in $D \geq 4$ dimensions, generalizing the one introduced in the $D = 4$ case \cite{dr10a,dr10b}. One can explicitly check that the first integral (\ref{eq:(2-5)quart}) and the positivity of energy condition (\ref{eq:(2-11)}) guarantee the monotonically decreasing character of the function $E(r)$ (for $Q > 0$), which must vanish asymptotically ($E(r \rightarrow \infty) = 0$). At $r = 0$ we can distinguish the cases where the field diverges at the center and those where it takes a finite value there. On the other hand, the positivity of the derivative of the Lagrangian function in Eq.(\ref{eq:(2-9)}) allows us to restrict the analysis to the case ($E>0, Q>0$) without loss of generality. This way we can assume polynomial-type behaviours for the ESS solutions around the center
\be
E(r \rightarrow 0,Q) \sim \nu_{1}(Q)r^{p} \ ,
\label{eq:(2-16)}
\en
and asymptotically
\be
E(r \rightarrow \infty,Q) \sim \nu_{2}(Q)r^{q} \ ,
\label{eq:(2-16)bis}
\en
where $\nu_{1}(Q)$ and $\nu_{2}(Q)$ are some $Q$-dependent constants, and the admissibility conditions constraint the values of the exponents to $p \leq 0$ and $q < 0$. At the center, $r=0$, the fields diverge for $p<0,$ while for $p=0$ they behave there as
\be
E(r \rightarrow 0,Q) \sim a - b(Q) r^{\sigma} \ ,
\label{eq:(2-17)}
\en
where the parameter $a$ (the maximum field strength) and the exponent $\sigma > 0$ are universal constants for a given model, whereas the coefficient $b(Q)$ is related to the charge of each particular solution as
\be
b(Q) Q^{\frac{\sigma}{D-2}} = \lim_{X \rightarrow a^{2}} (a - \sqrt{X})
\left[a \frac{\partial \varphi}{\partial X}\right]^{\frac{\sigma}{D-2}} = b_{0} \ ,
\label{eq:(2-18)}
\en
$b_{0} = b(Q = 1)$ being also a positive universal constant of the model.

\subsubsection{Asymptotic behaviour}

Let us consider first the asymptotic behaviour of the electric fields. In $D$ spacetime dimensions the generalized Coulomb field is the elementary solution of Maxwell electrodynamics (defined by $\varphi(X) \equiv X$) and its explicit form follows trivially from Eq.(\ref{eq:(2-5)quart}) as
\be
E(r,Q) = \frac{Q}{r^{D-2}} \ .
\label{eq:(2-19)}
\en
Starting from this expression, we shall distinguish the asymptotic cases for which the negative exponent $q$ in Eq.(\ref{eq:(2-16)bis}) is greater than, smaller than, or equal to $2-D$, corresponding to fields which are asymptotically damped slower than, faster than, or as the Coulomb field, respectively.  Moreover, the integral of energy for these spherically symmetric solutions, obtained from Eqs.(\ref{eq:(2-7)}) and (\ref{eq:(2-5)quart}), which reads
\be
\varepsilon = \omega_{(D-2)} \int_{0}^{\infty} dR \left(2 Q E - R^{D-2}\varphi \right) \ ,
\label{eq:(2-20)}
\en
converges asymptotically if $q < -1$ while diverges if $-1 \leq q < 0$. As a consequence we can classify the asymptotic behaviour in similar families as those found in $D=4$ dimensions. The InfraRed Divergent (IRD) cases, corresponding to $-1 \leq q < 0$, for which the fields are damped asymptotically but the integral of energy diverges at large $r$. The B1 cases, when $2-D < q < -1$, for which the fields are asymptotically damped slower than the Coulomb field and the integral of energy converges at large $r$. The B2 cases, corresponding to $q = 2 - D$, for which the fields are asymptotically Coulombian and the integral of energy converges at large $r$. Finally, the B3 cases, when $q < 2-D$, for which the fields are damped asymptotically faster than the Coulomb field and the integral of energy converges at large $r$.

\subsubsection{Central-field behaviour}

Similarly, let us classify the central-field behaviours. We can distinguish the cases with $p = 0$ and those with $p < 0$ in Eq.(\ref{eq:(2-16)}). When $p = 0$ the fields behave as in Eq.(\ref{eq:(2-17)}) around the center and the integral of energy (\ref{eq:(2-20)}) converges there. We shall denote this behaviour as cases A2, consistently with the conventions introduced in Refs.\cite{dr09,dr13} in $D=4$. For $p < 0$ the central fields diverge, but if $-1 < p <0$ the integral of energy converges there (cases A1). For $p \leq -1$ the fields and their integral of energy diverge at the center (UltraViolet Divergent or UVD cases). In summary, the families of models supporting finite-energy elementary solutions are, as in $D=4$, the combinations of those exhibiting simultaneously the A1 or A2 central-field behaviours, and the B1, B2 or B3 asymptotic behaviours, while any other combination implies divergent total energy.

\subsubsection{Behaviour of the Lagrangian density}

The behaviour (on vacuum and at large $X$) of the Lagrangian densities associated with these central and asymptotic behaviours of the elementary ESS solutions of UVD and A1 models is given by
\be
\varphi(X) \sim \alpha_{i} X^{\gamma_{i}} \ ,
\label{eq:(2-21)}
\en
where $\alpha_{i}$ and $\gamma_{i} (i=1,2)$ are positive constants which are related to the coefficients in Eqs.(\ref{eq:(2-16)}) or (\ref{eq:(2-16)bis}) via the first integral (\ref{eq:(2-5)quart}). Such relations between the coefficients and exponents as $r \sim 0$ ($X \rightarrow \infty$) read
\be
\nu_{1}(Q) = \left(\frac{\gamma_{1}\alpha_{1}}{Q}\right)^{\frac{p}{D-2}}\hspace{.1cm}; \hspace{.1cm} \gamma_{1} = \frac{1}{2} - \dfrac{D-2}{2p} \ .
\label{eq:(2-22)}
\en
For large $r \rightarrow \infty$ ($X \rightarrow 0)$, the corresponding relations are instead
\be
\nu_{2}(Q) = \left(\frac{\gamma_{2}\alpha_{2}}{Q}\right)^{\frac{q}{D-2}}\hspace{.1cm};
\hspace{.1cm} \hspace{.2cm} \gamma_{2} = \frac{1}{2} - \dfrac{D-2}{2q} \ .
\label{eq:(2-23)}
\en
In both cases the positivity of the energy condition, $\gamma_{i}>1/2$ (see Eq.(\ref{eq:(2-12)})), is fulfilled. The asymptotically Coulombian behaviour (\ref{eq:(2-19)}) corresponds to $\gamma_{2} = 1$.

\begin{figure}[t]
 \begin{center}
\includegraphics[width=0.45\textwidth]{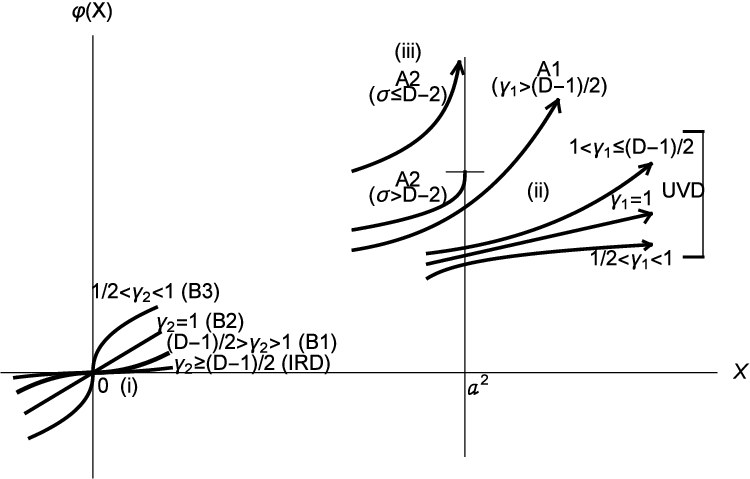}
\caption{Qualitative behaviour of the admissible Lagrangian densities $\varphi(X)$:0 (i) around the vacuum ($X \sim 0\hspace{0.1cm};\hspace{0.1cm} \varphi(X \sim 0) \sim X^{\gamma_{2}}$), corresponding to the three B cases and IRD asymptotic behaviours of the ESS solutions, (ii) for large ESS fields ($X \rightarrow \infty; \varphi(X \rightarrow \infty) \sim X^{\gamma_{1}}$), corresponding to the A1 and UVD central-field behaviours, and (iii) for finite maximum field-strength models ($X \leq a^{2} = E^{2}_{max}$), corresponding to the A2 central-field behaviour. The $\gamma_{i}$ constants are related to the central and asymptotic behaviours of the ESS fields through Eqs.(\ref{eq:(2-16)})-(\ref{eq:(2-16)bis}) and (\ref{eq:(2-21)})-(\ref{eq:(2-23)}). In the A2 cases (see Eq.(\ref{eq:(2-17)})) the Lagrangian density exhibits a vertical asymptote at $X = a^{2}$ (if $\sigma \leq D-2$) or takes a finite value with divergent slope there (if $\sigma > D-2$). In the intermediate range of $X>0$ values, matching the central and asymptotic regions, $\varphi(X)$ must be strictly monotonically increasing, for admissibility (see Eq.(\ref{eq:(2-12)})). This figure is qualitatively similar for any value of $D \geq 4.$}
\label{figure1}
 \end{center}
\end{figure}

In the A2 cases (finite central-fields) we have $p = 0$, and the Lagrangian densities behave around the center ($X = E^{2}(r=0) = a^{2}$) as
\be
\varphi(X) \sim \frac{2 \sigma b_{0}^{\frac{D-2}{\sigma}}}{D-2-\sigma}(a - \sqrt{X})^{\frac{\sigma-D+2}{\sigma}} + \Delta \ ,
\label{eq:(2-24)}
\en
if $\sigma \neq D-2.$ For models with $\sigma = D-2$ this behaviour does not depend explicitly on $D$, and is given by
\be
\varphi(X) \sim -2 b_{0} \ln(a - \sqrt{X}) + \Delta \ .
\label{eq:(2-25)}
\en
In these formulae the constants $\Delta$ depend on the value of $\sigma$. If $\sigma > D - 2$, then $\Delta = \varphi(X=a^{2})$, which are finite and universal constants for a given model. We see that in these cases the Lagrangian densities $\varphi(X)$ attain a finite value with divergent slope at $X = a^{2}$, i.e., at the maximum field strength. If $\sigma \leq D-2,$ the Lagrangian density exhibits a vertical asymptote on $X = a^{2}$. In these cases $\Delta$ can be calculated, after a straightforward procedure, once the explicit form of $\varphi(X)$ is given (see Ref. \cite{dr13} for details). This behaviour of the different admissible Lagrangian densities is plotted in Fig.\ref{figure1} for any $D \geq 4$ case.

\subsubsection{Behaviour of the energy function}

Once the classification of the admissible NEDs in $D-$dimensional spacetimes is given, let us analyze the behaviour of the energy for the associated elementary solutions. As already mentioned, for those solutions belonging to families B1, B2 and B3 the integral of energy converges asymptotically and the \textit{external energy function} (which is interpreted as the field energy contained outside the $D-2$ hypersphere of radius $r$) is defined as
\be
\varepsilon_{ex}(r,Q,D) = \omega_{(D-2)} \int_{r}^{\infty} dR \left(2 Q E - R^{D-2}\varphi \right) \ .
\label{eq:(2-26)}
\en
This function cannot be defined for elementary solutions of the models belonging to the IRD family, owing to the asymptotic divergence of this integral. In the same way, for models with central-field behaviour ESS solutions belonging to families A1 and A2, the energy integral is convergent around the center and the \textit{internal energy function} (the field energy contained inside the $D-2$ hypersphere of radius $r$) is defined as
\be
\varepsilon_{in}(r,Q,D) = \omega_{(D-2)} \int_{0}^{r} dR \left(2 Q E - R^{D-2}\varphi \right) \ .
\label{eq:(2-27)}
\en
Again, this function cannot be defined for elementary solutions of the UVD family models because it does not converge at the center in such cases. This way, for models belonging to combinations of A1 or A2 central-field behaviours and B1, B2 or B3 asymptotic behaviours, the total energy of the ESS solutions is finite and takes the form
\be
\varepsilon(Q,D) = \omega_{(D-2)} \int_{0}^{\infty} dR \left(2 Q E - R^{D-2}\varphi \right) \ .
\label{eq:(2-28)}
\en
For these six families supporting finite-energy ESS solutions we have the obvious relation
\be
\varepsilon(Q) = \varepsilon_{in}(\infty,Q) = \varepsilon_{ex}(0,Q) = \varepsilon_{in}(r,Q) + \varepsilon_{ex}(r,Q) \ .
\label{eq:(2-29)}
\en
When such finite-energy elementary solutions are linearly stable they are genuine non-topological solitons. The analysis of stability for such solitons has been performed in Ref. \cite{dr09} for the (flat) $D = 4$ case. The extension of such analysis to higher dimensions could be done in a similar way, but it lies beyond of the scope of this paper.

With more generality, we can define the field energy contained in the $(D-1)-$dimensional volume between two $(D-2)-$hyperspheres of radii $r_{1}$ and $r_{2}$ as

\be
\varepsilon(r_{1},r_{2},Q,D) = \omega_{(D-2)} \int_{r_{1}}^{r_{2}} dR \left(2 Q E - R^{D-2}\varphi \right).
\label{eq:(2-30)}
\en

The next step in our analysis is to determine the scale laws for the energy functions associated to the ESS solutions. They are obtained from Eq.(\ref{eq:(2-13)}) and the definitions (\ref{eq:(2-26)})-(\ref{eq:(2-28)}) for the energy integrals in the ESS cases, and read explicitly
\bea
\varepsilon(Q,D) &=& Q^{\frac{D-1}{D-2}} \varepsilon(Q = 1,D);
\nonumber\\
\varepsilon_{in}(r,Q,D) &=& Q^{\frac{D-1}{D-2}} \varepsilon_{in}\left(\frac{r}{Q^{\frac{1}{D-2}}},Q = 1,D \right);
\nonumber\\
\varepsilon_{ex}(r,Q,D) &=& Q^{\frac{D-1}{D-2}} \varepsilon_{ex}\left(\frac{r}{Q^{\frac{1}{D-2}}},Q = 1,D \right) \ ,
\label{eq:(2-31)}
\ena
or, equivalently, under the $\Gamma(\theta)$ group transformations
\bea
\varepsilon(\theta Q,D) &=& \theta^{\frac{D-1}{D-2}} \varepsilon(Q,D);
\nonumber\\
\varepsilon_{in}(\theta r,\theta^{D-2} Q,D) &=& \theta^{D-1} \varepsilon_{in}(r,Q,D);
\nonumber\\
\varepsilon_{ex}(\theta r,\theta^{D-2} Q,D) &=& \theta^{D-1} \varepsilon_{ex}(r,Q,D) \ ,
\label{eq:(2-32)}
\ena
where $\theta$ is a positive parameter.

The comparison of these results with those of Ref.\cite{dr13} for $D=4$ shows that the generalization of the analysis of NEDs in flat four-dimensional spacetime to the $D-$dimensional case does not introduce new essential qualitative features. In particular, the characterization of the different families of admissible models through the properties of their Lagrangian density functions remains qualitatively the same.

\subsection{Two illustrative examples}

\subsubsection{Born-Infeld}

Let us consider, as a first illustrative example, the generalization of the well known Born-Infeld model \cite{BI34} to $D$ spacetime dimensions. This model is defined by the Lagrangian density
\be
\varphi(X;\mu) = \frac{1 - \sqrt{1 - \mu^{2} X}}{\mu^{2}/2} \ .
\label{eq:(2-33)}
\en
where $\mu$ is a free parameter\footnote{Recently, it has been shown that compatibility of this NED in $D=4$ with hydrogen's ionization energy allows to constrain the Born-Infeld parameter as $\mu^{-1}>1.074 \times 10^{21}V/m$; see \cite{Akmansoy:2017vcy} for details.}.
In the limit $\mu \to 0 $, this function reduces to the Maxwell Lagrangian density, $\varphi(X) = X$. In addition, for small values of the field, $\mu^{2} X \ll 1$, it also approaches the Maxwell Lagrangian. Note that (\ref{eq:(2-33)}) is defined for $X \leq \mu^{-2}$ only, and $\varphi(X)$ exhibits at $X = \mu^{-2}$ an absolute maximum with divergent slope. Consistently with the classification introduced above, it belongs to the family A2 with $\sigma > D-2$ (see Eq.(\ref{eq:(2-24)})). Asymptotically it belongs to the family B2. The first integral (\ref{eq:(2-5)quart}) reads in this case
\be
r^{D-2} \varphi_X E = \frac{r^{D-2}  E}{\sqrt{1 - \mu^{2} E^{2}}} = Q \ ,
\label{eq:(2-34)}
\en
and leads to the explicit expression of the elementary electrostatic field
\be
E(r,\mu,Q,D) = \frac{Q}{\sqrt{r^{2(D-2)} + Q^{2} \mu^{2}}} \ .
\label{eq:(2-35)}
\en
As expected, this solution reduces to the Coulomb field (\ref{eq:(2-19)}) if $\mu = 0$ and behaves as this Coulomb field for large values of $r$:
\be
E(r \rightarrow \infty,\mu,Q,D) \sim \frac{Q}{r^{D-2}} \rightarrow 0 \ .
\label{eq:(2-36)}
\en
At the center the solution takes the finite value $E(0,\mu,Q,D) = 1/\mu$, in agreement with the A2 family properties, behaving at small $r$ as
\be
E(r \rightarrow 0,\mu,Q,D) \sim \frac{1}{\mu} \left(1 - \frac{r^{2(D-2)}}{2 Q^{2} \mu^{2}}\right) \rightarrow \frac{1}{\mu} \ ,
\label{eq:(2-37)}
\en
which  gives the characteristic parameters of the polynomial expansion of the field around the center through Eqs.(\ref{eq:(2-17)}) and (\ref{eq:(2-18)}):
\be
a = \frac{1}{\mu} \hspace{0.1cm}; \hspace{0.1cm} \sigma = 2(D-2)\hspace{0.1cm}; \hspace{0.1cm} b_{0} = b(Q) Q^{2} = \frac{1}{2\mu^{3}} \ .
\label{eq:(2-38)}
\en
Finally, the expression for the total energy of the elementary solutions can be obtained from Eqs.(\ref{eq:(2-28)}), (\ref{eq:(2-33)}) and (\ref{eq:(2-35)}) and reads
\be
\varepsilon(\mu,Q,D) = \frac{2\omega_{(D-2)}}{\mu^{\frac{D-3}{D-2}}} Q^{\frac{D-1}{D-2}} I(D) \ ,
\label{eq:(2-39)}
\en
where the integral
\bea
I(D) &=& \int_{0}^{\infty} dy \left( \sqrt{y^{2(D-2)} + 1} - y^{(D-2)}\right),
\label{eq:(2-40)}
\ena
yields a finite contribution provided that $D \geq 4$. In the limit $\mu \rightarrow 0$ the energy (\ref{eq:(2-39)}) diverges. This is consistent with the fact that the model becomes the linear Maxwell electrodynamics in this limit, and its ESS solutions become the energy-divergent Coulomb field.

\subsubsection{Euler-Heisenberg and its extensions}

As a second illustrative example let us consider a $D-$dimensional version of the Euler-Heisenberg model,  defined in $D=4$ by the Lagrangian density
\be
\varphi(X;\lambda) = X + \lambda X^{2} \ ,
\label{eq:(2-41)}
\en
where the parameter $\lambda(>0)$ gives the strength of the nonlinear coupling. This model satisfies the admissibility conditions and reduces to the Maxwell Lagrangian in the limit $\lambda \rightarrow 0$. The ESS solutions in this case are obtained from Eq.(\ref{eq:(2-5)quart}), which now takes the form
\be
2\lambda E^{3} + E(r,q) = \dfrac{Q}{r^{D-2}} \ ,
\label{eq:(2-42)}
\en
and can be solved explicitly through the Tartaglia formula, leading to
\be
E(r,Q) = \left[\dfrac{v}{r^{D-2}} + \sqrt{\Delta}\right]^{1/3} + \left[\dfrac{v}{r^{D-2}} - \sqrt{\Delta}\right]^{1/3} \ ,
\label{eq:(2-43)}
\en
where $u = \frac{2}{3\lambda}$, $v= \frac{2Q}{\lambda}$ and $\Delta = v^{2}/r^{2(D-2)} + u^{3} > 0$. Near the center these fields diverge as $E(r \rightarrow 0,Q) \sim r^{-(D-2)/3}$, while they are asymptotically Coulombian: $E(r \rightarrow \infty,Q) \sim r^{-(D-2)}$ (case B2).

The central-field behaviour ($p=-(D-2)/3$), together with the admissibility conditions, endorse the decreasing and concave character of the exterior integral of energy, which converges as $r \rightarrow 0$ in $D=4$ dimensions (A1 cases). For $D \geq 5$ the energy of the elementary solutions diverges and the Euler-Heisenberg model in these dimensions belongs to the UVD family. The expression for the finite total energy of the elementary solutions in $D=4$ can be obtained from Eqs.(\ref{eq:(2-28)}), (\ref{eq:(2-41)}) and (\ref{eq:(2-43)}) and reads:
\bea
\varepsilon(Q,D=4) &=& \frac{16\pi Q^{3/2}}{3} \int_{0}^{\infty} \frac{dy}{\sqrt{y(1 + 2\lambda y^{2})}} \nonumber \\
&=& \frac{8\pi Q^{3/2}}{3\lambda^{1/4}} B\left(\frac{1}{4},\frac{1}{4}\right) \ ,
\label{eq:(2-44)}
\ena
where $B(x,y) = \int_{0}^{1} t^{x-1}(1-t)^{y-1} dt$ is the Euler integral of first kind.

The model (\ref{eq:(2-41)}) can be naturally generalized to the polynomial form \cite{dr09,DRG15}
\be
\varphi(X,\lambda_{i}) = X + \sum_{i=2}^{N}\lambda_{i} X^{i} \ ,
\label{eq:(2-45)}
\en
which is defined by the $(N-1)$ parameters $\lambda_{i}$. With a proper choice of these parameters, this model (in $D=4$) corresponds to an effective Lagrangian of quantum electrodynamics accounting for the higher-order contributions to the photon propagation of the vacuum polarization in the perturbative expansion \cite{EHQED1,EHQED2}. In $D>4$ dimensions these models have finite-energy elementary solutions if
\be
N > \frac{D-1}{2} \ ,
\label{eq:(2-46)}
\en
and belong to the central field A1 family. Otherwise they are UVD models.

\section{Gravitating nonlinear electrodynamics in $D \geq 4$ spacetime dimensions} \label{sectionIII}

We shall consider now the interaction of NED fields with gravitational fields, assuming a minimal coupling and a cosmological constant term. The action describing such dynamical systems is given by
\be
\mathcal{S}=\mathcal{S}_G+\mathcal{S}_{NED} = \int d^D x \sqrt { \vert g \vert } \left[ \frac { R-(D-2)\Lambda}{ 2\chi } - \varphi (X) \right] \ ,
\label{eq:(3-1)}
\en
where $\vert g \vert$ is the determinant of the metric tensor $g_{\mu\nu}$, $\Lambda$ is the cosmological constant, and $\chi$ is related to the $D$-dimensional gravitational Newton's constant, $G_{D}$. As usual, the extremum condition of this action under the variation of the matter fields leads to the electromagnetic field equations, which generalize (\ref{eq:(2-2)}) to curved space as
\be
\nabla_{\mu} [\varphi_X F^{\mu \nu}] = 0 \ ,
\label{eq:(3-2)}
\en
while variation with respect to the metric tensor yields the Einstein equations
\bea
G_{\mu \nu} + \frac{D-2}{2} \Lambda g_{\mu \nu }&=& R_{\mu \nu} - \frac{1}{2} \left[ R - (D-2)\Lambda\right] g_{\mu \nu }
\nonumber \\ &=& -\chi T_{\mu \nu} \ ,
\label{eq:(3-3)}
\ena
where the symmetric form of the electromagnetic energy-momentum tensor is given by
\be
T_{\mu \nu} = 2 \varphi_X F_{\mu \beta} {F^\beta}_{\nu} - \varphi g_{\mu \nu} \ .
\label{eq:(3-4)}
\en

\subsection{The elementary solutions}

Looking for (electro) static spherically symmetric (elementary) solutions of Einstein's equations, a general coordinate system suitably adapted to these symmetries can be characterized by the line element
\be
{ds}^{2} = \lambda (r){dt}^{2} - \frac {{dr}^{2}}{\mu(r)} - {r}^{2}{d\Omega}_{D-2}^{2} \ ,
\label{eq:(3-5)}
\en
where the angular contribution is the metric on the $S^{D-2}$ sphere and takes the form
\be
{ d\Omega  }_{ D-2 }^{ 2 }={ { d\theta  }_{ 1 }^{ 2 } } +
\Sigma_{i=2}^{D-2} \prod_{ j=1 }^{ i-1 }{ { \sin }^{ 2 }{ \theta  }_{ j } } { d\theta  }_{ i }^{ 2 } \ .
\label{eq:(3-6)}
\en
In this coordinate system the metric tensor is diagonal and the only non-vanishing components of the electrostatic field tensor are $F_{01} = -F_{10} \equiv E(r)$. As a consequence, Eq.(\ref{eq:(3-4)}) leads to the following expressions for the nonvanishing components of the ESS energy-momentum tensor
\bea
{T_0}^{0} &=& {T_1}^{1} = 2X \varphi_X - \varphi(X), \nonumber\\
{T_i}^{i} &=& -\varphi(X) \hspace{.3cm} (i \geq 2) \ ,
\label{eq:(3-7)}
\ena
which hold when $X = E^{2}(r)$. As in the $D=4$ cases, these relations will lead to a simplification in the form of the line element (\ref{eq:(3-5)}). Indeed, let us obtain the explicit form of the Einstein equations (including the cosmological term) with static, spherical symmetry. This calculation is standard and has been done in the literature for many particular models. The extension to the case of general G-NEDs in $D$ dimensions is straightforward (see the Appendix). Using the first of Eqs.(\ref{eq:(3-7)}) and the expressions (\ref{eq:(A-1)}) and (\ref{eq:(A-2)}) of the components of the Einstein tensor given in the Appendix, the subtraction of the $({_0}^0)$ and $({_1}^1)$ components of the Einstein equations leads to
\be
\frac{d}{dr} \left( \sqrt { \frac { \lambda  }{ \mu  }}\right) = 0 \ .
\label{eq:(3-12)}
\en
Upon redefinition of the time coordinate this equation can be integrated, without loss of generality, as
\be
\mu(r) = \lambda(r) \equiv g(r) \ ,
\label{eq:(3-13)}
\en
where we have introduced the function $g(r) = g_{00}(r)$. This way, the line element (\ref{eq:(3-5)}) gets simplified and takes the Schwarzschild-like form
\be
{ds}^{2} = g(r){dt}^{2} - \frac {{dr}^{2}}{g(r)} - {r}^{2}{d\Omega}_{D-2}^{2} \ .
\label{eq:(3-14)}
\en
We see that the determinant of the metric tensor in these Schwarzschild-like coordinates has the same form as the determinant of the flat spacetime metric tensor in spherical coordinates. Consequently, in writing explicitly the expression of the electrostatic field equations (\ref{eq:(3-2)}) in the ESS cases we have
\be
\frac{d}{dr} \left[ \sqrt{\vert g \vert} \varphi_X E(r)\right] = 0 \ .
\label{eq:(3-14)bis}
\en
The form of the metric determinant is written as
\be
\vert g \vert = -r^{2(D-2)} \Theta( \theta_{i} ) \ ,
\label{eq:(3-14)ter}
\en
where $\Theta(\theta_{i})$ contains the angular dependence only. Thus, Eq.(\ref{eq:(3-14)bis}) can be integrated leading to a first integral having the same form (\ref{eq:(2-5)quart}) as in the flat spacetime. We conclude that the expression of the ESS field associated to a given G-NED, as a function of the radial coordinate of the Schwarzschild-like coordinate system (\ref{eq:(3-14)}), is the same as that of the elementary solution of the same NED in flat space as a function of the radial coordinate of the polar coordinate system. This is a key element in order to generalize to curved spacetimes all the results obtained for NEDs in flat spacetimes in any $D\geq4$ dimensions.

Let us come back now to the integral (\ref{eq:(A-9)}) of the Einstein equations obtained in the Appendix. Taking the limit $r_{2} \rightarrow \infty$ and identifying $r_{1} = r$ we obtain
\be
g(r,M,Q,\Lambda,D) = 1 - \frac{2M}{r^{D-3}} - \frac{\Lambda r^{2}}{D-1} + \frac{2 \varepsilon_{ex}(r,Q,D)}{r^{D-3}} \ ,
\label{eq:(3-20)}
\en
where we have defined the integration constant $M$ as
\be
M = - \frac{1}{2} \lim_{r \rightarrow \infty} \left[ r^{D-3} (g(r) - 1) + \frac{\Lambda r^{D-1}}{D-1}\right] \
\label{eq:(3-21)}
\en
and fixed the gravitational constant $\chi$ as
\be
\chi = (D-2)\omega_{(D-2)} \ ,
\label{eq:(3-21)bis}
\en
which is tantamount to set units $G_{D} = c = 1$. The constant $M$ plays the role of a mass parameter, which is related to the ADM mass through \cite{myers86}:
\be
M = \dfrac{8\pi M_{ADM}}{(D-2)\omega_{(D-2)}} \ .
\label{eq:(3-21)ter}
\en
The line element (\ref{eq:(3-14)}) with the metric function (\ref{eq:(3-20)}) contains several interesting limit cases:

\begin{enumerate}
  \item In absence of the cosmological term ($\Lambda = 0$) and of electrostatic field ($Q = 0$) it becomes the well known generalization to $D$ dimensions of the Schwarzschild gravitational field:
\be
g(r,M,Q=0,\Lambda=0,D) = 1 - \frac{2M}{r^{D-3}} \ .
\label{eq:(3-22)}
\en

  \item With $\Lambda = 0$ it becomes the metric of asymptotically flat charged BH configurations associated to admissible G-NEDs, whose external energy function is given by Eq.(\ref{eq:(2-26)}):
\be
g(r,M,Q,\Lambda=0,D) = 1 - \frac{2M}{r^{D-3}}+ \frac{2 \varepsilon_{ex}(r,Q,D)}{r^{D-3}} \ ,
\label{eq:(3-23)}
\en
which generalize to $D>4$ dimensions the gravitating ESS solutions in $D=4$, discussed in Ref.\cite{dr13}.

  \item If $Q = M = 0$ the metric becomes the de-Sitter (if $\Lambda > 0$) or AdS
  (if $\Lambda < 0$) spaces in $D$ dimensions:
\be
g(r,M=0,Q=0,\Lambda,D) = 1 - \frac{\Lambda r^{2}}{D-1} \ .
\label{eq:(3-24)}
\en
  \item With $Q = 0$ it becomes the generalization to $D$ dimensions of the Kottler-Weyl spacetime in $D=4$, representing a Schwarzschild-like BH embedded in de-Sitter or AdS spaces, depending on the sign of $\Lambda$:
\be
g(r,M,Q=0,\Lambda,D) = 1 - \frac{2M}{r^{D-3}} - \frac{\Lambda r^{2}}{D-1} \ .
\label{eq:(3-25)}
\en

  \item Finally, the full metric (\ref{eq:(3-20)}) describes asymptotically de-Sitter or AdS charged BHs associated to admissible G-NEDs. In particular, if the source is the Maxwell electrodynamics, the metric describes asymptotically de-Sitter or AdS Reissner-Nordstr\"om BHs in $D$ dimensions.

\end{enumerate}

It is worth pointing out that the ESS black holes resulting from the metric function (\ref{eq:(3-20)}) are not the only possible ones in $D > 4$ dimensions. Indeed, in such cases there exist also \textit{topological} BHs, characterized by the topology of their event horizons, which can be negative or zero curvature hypersurfaces \cite{vanzo}. For these topological BHs the metric function reads
\be
g(r,M,Q,\Lambda,D) = k - \frac{2M}{r^{D-3}} - \frac{\Lambda r^{2}}{D-1} + \frac{2 \varepsilon_{ex}(r,Q,D)}{r^{D-3}}  \ ,
\label{eq:(3-26)}
\en
where the constant $k$ can take the values $k = 0$ (zero curvature event horizon) or $k = -1$ (negative curvature event horizon), besides the value $k=1$ in Eq.(\ref{eq:(3-20)}), for which the event horizon exhibits the usual spherical topology. It has been shown in several particular examples that different horizon topologies lead to quite different behaviours of the corresponding BH solutions \cite{Birmingham:1998nr,lemos1}. This topic is of great interest, deserving an in-depth analysis which lies beyond the scope of this work, where we shall be concerned with topologically spherical horizon cases only.

\subsection{Asymptotically flat black holes}

Higher-dimensional ($D>4$) black holes, associated to G-NEDs, have been studied in the literature only for a few models, with particular emphasis in the Reissner-Nordstr\"om and Born-Infeld ones \cite{NEDads1,NEDads2,NEDads3,Peca:1998cs,Hendihig}. However, the general analysis of this issue, containing the full set of \textit{admissible} NEDs, is possible by using the properties described in Section \ref{sectionII}, characterizing the different families of NEDs in flat spacetime. This will allow us to determine the features of the BHs associated to a given model just by inspecting the functional form of its Lagrangian density.

Let us start in this section by considering the asymptotically flat ($\Lambda=0$) cases. The metric functions in these cases take the form (\ref{eq:(3-23)}) and their behaviour as a function of $r$ is governed by that of $\varepsilon_{ex}(r,Q,D)$. The first derivative of this function takes the form
\bea
\label{eq:(3-27)}
\frac{d}{dr} \varepsilon_{ex}(r,Q,D) &=& -\omega_{(D-2)} \left(2 Q E - r^{D-2}\varphi \right) \\
&=& -\omega_{(D-2)} {T_0}^{0} < 0  \ , \nonumber
\ena
the last inequality resulting from the constraint on the positivity of the energy. This simply means that the external energy is a decreasing function. Moreover, the second derivative of $\varepsilon_{ex}$ can be calculated from (\ref{eq:(3-27)}) and the first integral (\ref{eq:(2-5)quart}), and reads
\be
\frac{d^{2}}{dr^{2}} \varepsilon_{ex}(r,Q,D) = (D-2)\omega_{(D-2)}r^{D-3}\varphi > 0  \ ,
\label{eq:(3-28)}
\en
which means that $\varepsilon_{ex}(r,Q,D)$ is a monotonically decreasing and \textit{concave} function. The behaviour of this function at small and large $r$ depends on the particular family. For large $r$ this function vanishes in cases B1, B2 and B3. Consequently, in Eq.(\ref{eq:(3-23)}) the $r-$dependent dominant term, as $r \rightarrow \infty,$ is $-2M/r^{D-3}$, and the metric $g(r)$ approaches asymptotic flatness as the Schwarzschild solution. As in the $D=4$ case \cite{dr10a,dr10b}, we shall call this behaviour as ``asymptotically normal" in the $D > 4$ cases. At the center $\varepsilon_{ex}$ converges for the families A1 and A2 and exhibits a vertical asymptote there in the UVD cases.

For the IRD families, the external energy function is not well defined and Eq.(\ref{eq:(3-23)}) makes no sense. In these cases we must integrate the Einstein equations in terms of the internal energy function $\varepsilon_{in}(r,Q,D)$ in Eq.(\ref{eq:(2-27)}), which is well defined when the central behaviour belongs to families A1 and A2. This integration leads to
\be
g(r,M,Q,\Lambda=0,D) = 1 + \frac{C}{r^{D-3}} - \frac{2 \varepsilon_{in}(r,Q,D)}{r^{D-3}}  \ ,
\label{eq:(3-29)}
\en
where $C$ is an integration constant and $\varepsilon_{in}$ diverges at large $r$ slower than $r^{D-3}$. Consequently, the last term is dominant in this equation and $g(r)$ approaches asymptotic flatness at large $r$, but slower than the Schwarzschild field (``asymptotically anomalous" behaviour). The models belonging to the UVD-IRD families can also be treated by the same methods as in the $D=4$ case \cite{dr10b} and exhibit also asymptotically anomalous behaviours. In what follows we shall discuss just the models with asymptotically normal behaviour, the only ones for which the thermodynamic analysis of their BH solutions can be consistently carried out.

In looking for the horizons of the gravitating ESS configurations we must find the zeroes of $g(r)$. From (\ref{eq:(3-23)}) the condition $g(r_{h}) = 0$ leads to the relation
\be
M(r_{h},Q,D) = \frac{r_{h}^{D-3}}{2} + \varepsilon_{ex}(r_{h},Q,D) \ ,
\label{eq:(3-30)}
\en
where $r_{h}$ is the horizon radius. Moreover, from the definition (\ref{eq:(3-30)}) and the scale law (\ref{eq:(2-31)}) for the external energy, we can obtain the corresponding scale law for the mass as a function of $r_{h}$ and $Q$:
\bea
M(r_{h},Q,D) &=& \frac{1}{2} \left(1-Q^{\frac{2}{D-2}}\right) Q^{\frac{D-3}{D-2}}R_{h}^{D-3} \nonumber\\
&+& Q^{\frac{D-1}{D-2}} M(R_{h},Q=1,D)  \ ,
\label{eq:(3-30)bis}
\ena
where $R_{h} = r_{h}/Q^{\frac{1}{D-2}}$ is the normalized horizon radius. Alternatively, we can write the scale law of the mass in terms of the transformations (\ref{eq:(2-15)}) by using Eq.(\ref{eq:(2-32)}) as
\bea
M(\theta r_{h},\theta^{D-2}Q,D) &=& \theta^{D-1} M(r_{h},Q,D)  \nonumber \\
&+& \frac{\theta^{D-3}(1-\theta^{2})}{2} r_{h}^{D-3} \
\label{eq:(3-30)ter}
\ena
and it is straightforward to verify the group representation character of these transformations. These formulae generalize to $D>4$ the expressions already obtained in $D=4$ dimensions \cite{dr13}.

\subsection{Extreme black holes and other configurations}

The $M-r_h$ relation (\ref{eq:(3-30)}) can be analyzed for the different families in $D>4$, taking into account the generic behaviour of the corresponding external energy function (\ref{eq:(2-26)}) defined by equations (\ref{eq:(3-27)}) and (\ref{eq:(3-28)}). In a similar way as in the $D=4$ case, this analysis gives, in particular, the horizon structure of the associated BH solutions.

\begin{figure}[h]
 \begin{center}
\includegraphics[width=0.45\textwidth]{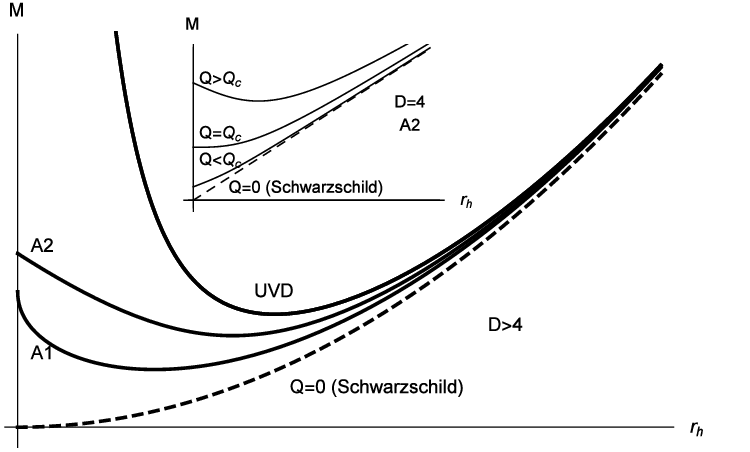}
\caption{Qualitative $M-r_h$ diagram for the asymptotically Schwarzschild BH solutions of admissible G-NEDs belonging to families with central-field behaviours A1, A2, and UVD in $D>4$ dimensions. The curves correspond to fixed values of $Q$. There are always unique minima in these curves corresponding to \textit{extreme black holes}. Naked singularities correspond to the configurations with a mass below the value of that of the extreme BH for a given charge, and exist in all cases. Furthermore, all families support \textit{two-horizon BHs}. The A1 and A2 families, supporting soliton solutions in flat space, exhibit also \emph{non-extremal single-horizon BH} solutions for values of $M$ above the total electromagnetic energy of the configuration ($M>\varepsilon(Q,D)$). The increasing parts of these curves correspond to the external event horizons, whose radii increase monotonically with the mass. The dashed curve, to which all the curves converge at large $r_{h}$, corresponds to the $M-r_{h}$ relation for the Schwarzschild BHs. The small frame displays the qualitative behaviour for A2 family in $D=4$ spacetime dimensions, where extreme and non-extreme black points arise for $Q = Q_{c}$ and $Q < Q_{c}$, respectively, $Q_{c} = (16\pi a)^{-1}$ being the critical value of the charge in these cases.}
\label{2}
 \end{center}
\end{figure}

If we look for the extrema of the mass parameter as a function of $r_{h}$ we must search for the zeroes of the derivative of Eq.(\ref{eq:(3-30)}), which reads
\be
\frac{\partial M}{\partial r_{h}}\Big \vert_{Q} = \frac{(D-3)}{2} r_{h}^{D-4} -\omega_{(D-2)} (2 Q E - r_{h}^{D-2} \varphi) = 0 \ .
\label{eq:(3-31)}
\en
From the monotonically decreasing and concave character of $\varepsilon_{ex}(r_{h},Q,D)$ it is obvious that there is an unique solution of this equation for every value of $Q$, which corresponds to a minimum $M_{e}(Q,D)$ of the mass-radius curve (see Fig.\ref{2}). The horizontal straight lines corresponding to different values of the mass parameter $M$ cut the curves associated to different constant values of the charge $Q$, and we see that there may be zero, one or two cut points for each of these lines. Such cut points define horizons of the ESS black hole configurations. We see that we can have configurations with, \textit{at most}, two horizons: one internal Cauchy horizon and one external event horizon. If the value of the mass parameter corresponds to the minimum ($M_{e}$) of a fixed-charge curve, defined by a simultaneous solution of both Eqs.(\ref{eq:(3-30)}) and (\ref{eq:(3-31)}), we have a \textit{extreme black hole} configuration, with an unique degenerate horizon. For values of the mass parameter below the value $M_{e}(Q,D)$ of the extreme BH (for a given charge) there are not horizons and the corresponding solutions are \textit{naked singularities}. For the families for which the total electrostatic energy of the ESS solutions $\varepsilon(Q,D)$ is finite (A1 and A2 families) there are also single horizon non-extreme BH configurations, for which the mass parameter exceeds the soliton energy in flat space ($M > \varepsilon(Q,D)$).

Let us point out an important feature arising in $D=4$ spacetime dimensions: The slopes of the curves $M(r_{h})$ around $r_{h} \sim 0$ are strictly negative for $D>4$ (negative-finite in A2 cases and negative-divergent in A1 and UVD cases, for $r_{h} \rightarrow 0$). However, in $D=4$ for the A2 cases, Eq.(\ref{eq:(3-31)}) becomes (see Eq.(\ref{eq:(2-17)}))
\be
\frac{\partial M}{\partial r_{h}}\Big \vert_{Q} \sim \frac{(D-3)}{2} r_{h}^{D-4} - 2\omega_{(D-2)} Qa \ .
\label{eq:(3-32)}
\en
We see that, in $D=4$ dimensions, there are $M(r_{h},Q_{c})$ curves associated with a critical value of the charge ($Q_{c} = (16\pi a)^{-1}$) whose slope vanishes at $r_{h} = 0$, defining in this way extreme ``black point" configurations \cite{dr13} (there are also non-extreme black points if $Q<Q_{c}$). It is obvious that no such configurations can exist for admissible G-NEDs if $D>4$ and we conclude that these black point configurations are uniquely ascribed to A2 models in four spacetime dimensions.

It is thus clear that the number of horizons of the BH solutions is mainly governed by the central-field behaviour of the elementary solutions associated to the A1, A2 or UVD families, no matter their asymptotic behaviour\footnote{As in the $D=4$ case \cite{dr10b}, a similar analysis can be performed for the asymptotically anomalous BHs resulting from IRD families and leading to a similar horizon structure. As already mentioned, such BHs do not exhibit consistent thermodynamic properties and will not be further discussed here.}. Thus, we conclude that the charged elementary solutions of Einstein equations minimally coupled to physically admissible NEDs in $D>4$ spacetime dimensions are necessarily asymptotically Schwarzschild or anomalous two-horizon BHs, single-horizon (extreme or non-extreme) BHs, or naked singularities. The existence of extreme and non-extreme black points is an unique feature of the admissible A2 gravitating NEDs in $D=4$ spacetime dimensions.

The set of extreme BHs can be characterized from the relation $r_{he}(Q)$ between the horizon radius and the charge of these objects, which is implicit in Eq.(\ref{eq:(3-31)}). Using the first integral (\ref{eq:(2-5)quart}) this equation boils down to
\be
Q^{\frac{2}{D-2}} = \left( \frac{D-3}{2\omega_{(D-2)}}\right) \left( \dfrac{ R_{he}^{D-4}}{2E_{he} - R_{he}^{D-2} \varphi(E_{he}^2)} \right) \ ,
\label{eq:(3-32)bis}
\en
where $E_{he}$ is the strength of the electric field on the horizon and $R_{he} = r_{he}/Q^{\frac{1}{D-2}}$ is the normalized extreme horizon radius. Once the explicit expression of the Lagrangian density is specified this equation takes the form $Q=f(R_{he})$, owing to the scale law (\ref{eq:(2-13)}) of the electric field, and leads to the relation between $Q$ and $r_{he}$ for extreme BHs in a direct way.

\subsection{Comparison with asymptotically AdS black holes}

Let us now consider the case of coupling of the Einstein-Hilbert action with a cosmological term to admissible NEDs in $D\geq4$ spacetime dimensions. The metric function for gravitating ESS solutions with topologically spherical horizons is defined by Eq.(\ref{eq:(3-20)}). As already mentioned, here we consider the asymptotically AdS ($\Lambda < 0$) cases only\footnote{The de-Sitter case introduces additional elements which require an extended analysis going well beyond the scope of this work. For studies of some particular NEDs in the de-Sitter backgrounds see e.g. \cite{NEDads1,NEDads2,NEDads3}.}. Let us write the metric function in this case as
\be
g(r,M,Q,l,D) = 1 - \frac{2M}{r^{D-3}} + \frac{r^{2}}{l^{2}} + \frac{2 \varepsilon_{ex}(r,Q,D)}{r^{D-3}} \ ,
\label{eq:(3-33)}
\en
where we have defined the constant
\be
l = \sqrt{\frac{1-D}{\Lambda}},
\label{eq:(3-34)}
\en
which has dimension of length and characterizes the AdS spacetime. In the metric function (\ref{eq:(3-33)}) the cosmological term dominates at large $r$. Therefore it describes a elementary gravitational field which reaches asymptotically the AdS metric and is characterized by the three parameters: $M$, $Q$, and $l$.

The number of horizons of these asymptotically AdS black holes in $D \geq 4$ is the same as that of the asymptotically flat BHs analyzed previously. Indeed, to obtain the location of the horizons we must solve the equation $g(r_{h}) = 0,$ which now reads
\be
M_{AdS} = \frac{r_{h}^{D-3}}{2} \left(1 + \frac{r_{h}^{2}}{l^{2}} \right) + \varepsilon_{ex}(r_{h},Q,D) \ ,
\label{eq:(3-35)}
\en
where the meaning of the index $AdS$ is obvious. This equation must be compared to Eq.(\ref{eq:(3-30)}). The first terms on the right-hand side in both equations are monotonically increasing and concave parabolic branches. They vanish (as well as their derivatives) at $r_{h}=0,$ exhibiting similar behaviours there. Although both terms increase with $r_{h}$ at different rates, the mass-$r_{h}$ relations behave qualitatively in a similar way. Consequently, the qualitative diagram of Fig.\ref{2} is also similar in both cases $\Lambda \leq 0$ and we conclude that the numbers of horizons of the different classes of BHs in both scenarios are the same.

If we look now for the scale law of the mass as a function of $r_{h}$ and $Q$ for fixed $l$, following the same steps as in the derivation of Eqs.(\ref{eq:(3-30)bis}) and (\ref{eq:(3-30)ter}), the cosmological term disappears from the final explicit expressions, which are the same in both asymptotically flat and AdS cases. This is a consequence of the fact that the underlying scale symmetries come from the NED sector, which is common to both cases and independent of the cosmological term.

\begin{figure}[h]
 \begin{center}
\includegraphics[width=0.45\textwidth]{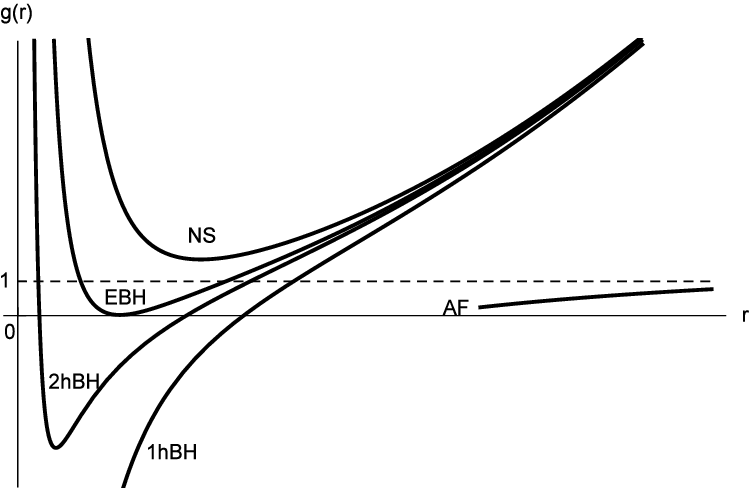}
\caption{The metric function $g(r)$ for ESS black holes embedded in AdS spacetime. At short distances, $g(r \rightarrow 0)$ diverges to $\pm \infty,$ depending on the family and the range of parameters (see the main text). At large $r$ the metric function reaches the parabola $1 + r^{2}/l^{2}$ (asymptotically AdS behaviour). In the intermediate region the different configurations (naked singularities, two-horizon BHs, extreme and non-extreme one-horizon BHs) follow from the cut points of the curves with the horizontal axis, which define the location of the horizons. The AF curve displays the large $r$ behaviour of the asymptotically flat BHs.}
\label{figure3}
\end{center}
\end{figure}

Obviously, the form of the metric function $g(r,Q,D)$ for large $r$ goes to one in the asymptotically flat cases, and diverges as
\be
g(r \rightarrow \infty,Q,l,D) \sim 1 + \frac{r^{2}}{l^{2}} \ ,
\label{eq:(3-36)}
\en
in the asymptotically AdS cases (see Fig.\ref{figure3}). As can be seen from Eqs.(\ref{eq:(2-16)}) and (\ref{eq:(2-30)}), at small $r$ (both in asymptotically flat and asymptotically AdS cases) the metric function behaves as
\bea
\label{eq:(3-37)}
g(r \rightarrow 0,Q,D) &\sim& \frac{2(\varepsilon(Q,D) - M)}{r^{D-3}}  \\
&-& \frac{4 Q \nu_{1}(Q) (D-2)
\omega_{(D-2)}}{(D-2-p)(p+1)}\dfrac{r^{p+1}}{r^{D-3}} \rightarrow \pm \infty \ ,  \nonumber
\ena
in the A1 cases; as
\be
g(r \rightarrow 0,Q,D) \sim \frac{2(\varepsilon(Q,D) - M)}{r^{D-3}} -  \frac{4 \omega_{(D-2)}Qa}{r^{D-4}} \rightarrow \pm \infty \ ,
\label{eq:(3-38)}
\en
in the A2 cases; and as
\be
g(r \rightarrow 0,Q,D) \sim \frac{2\varepsilon_{ex}(r \rightarrow 0,Q,D)}{r^{D-3}} \rightarrow +\infty \ ,
\label{eq:(3-39)}
\en
in the UVD cases. We see that in the finite-energy cases ($\varepsilon(Q,D) = \varepsilon_{ex}(r=0,Q,D) < \infty$) the metric function $g(r)$ diverges at the center to $\mp \infty$, depending on the sign of $M - \varepsilon(Q,D)$. If $M = \varepsilon(Q,D)$, then $g(r)$ diverges at the center to $-\infty$ in A1 and A2 cases (excepting in $D=4$, where $g(0,Q,D=4)$ can reach a finite value \cite{dr10a}). In the UVD cases we have always $g(r \rightarrow 0,Q,D) \rightarrow +\infty$.

Concerning the extreme BHs in these AdS cases, the equation relating the charge $Q$ and the normalized extreme horizon radius $R_{he}$ (the counterpart of Eq.(\ref{eq:(3-32)bis}) of the asymptotically flat cases) can be obtained in a similar way from Eq.(\ref{eq:(3-35)}), and reads
\be
Q^{\frac{2}{D-2}} = \frac{D-3}{2\omega_{(D-2)}}\left( \dfrac{ R_{he}^{D-4}}{2E_{he} - R_{he}^{D-2} \left(\varphi(E_{he}^2) + \frac{(D-1)}{2\omega_{(D-2)} l^{2}}\right)} \right) \ .
\label{eq:(3-40)}
\en
where a new term containing the cosmological length $l^2$ has been picked up.

\section{Thermodynamics of asymptotically flat and asymptotically AdS black holes} \label{sectionIV}

In this section we shall introduce the thermodynamic problem for the charged non-rotating, asymptotically Schwarzschild and AdS black hole solutions of the Einstein equations minimally coupled to admissible NEDs in $D\geq4$ spacetime dimensions, generalizing the results already obtained in $D=4$ \cite{dr13} to the present scenarios\footnote{As we shall see, for asymptotically AdS black holes a consistent thermodynamic analysis makes sense only if the underlying NEDs belong to the asymptotically B-cases.}. Moreover, regarding the issue of the scale transformations of the thermodynamic functions, we shall go beyond the analysis of \cite{dr13} by exploiting some simple consequences of their group structure. As mentioned in the introduction, this step, besides the already obtained extreme BH expressions (\ref{eq:(3-32)bis}) and (\ref{eq:(3-40)}), will allow us (in section \ref{sectionV}) for a large improvement in the analysis of the thermodynamic properties.

\subsection{Thermodynamics of asymptotically flat black holes in $D$ dimensions}

Let us come back to Eq.(\ref{eq:(3-30)}), which gives the mass-$r_{h}$ relation for asymptotically flat BHs. Differentiating this equation with respect to $r_{h}$ and $Q$ we obtain the expression
\be
dM = \frac{\partial M}{\partial r_{h}}\Big \vert_{Q} d r_{h} + \frac{\partial M}{\partial Q}\Big \vert_{r_{h}} d Q  \ ,
\label{eq:(4-1)}
\en
which will lead us to the explicit form of the first law of BH thermodynamics in $D$ dimensions. Indeed, let us first rewrite the expression (\ref{eq:(3-31)}) of the derivative of $M$ with respect to $r_{h}$ under the form:
\be
\frac{\partial M}{\partial r_{h}}\Big \vert_{Q} = \frac{(D-3)}{2} r_{h}^{D-4} - \omega_{D-2} r_{h}^{D-2} {T_0}^{0}  \ .
\label{eq:(4-2)}
\en
Assume now the usual definition of the \textit{entropy} as the fourth of the horizon area, i.e.:
\be
S = \frac{\omega_{(D-2)} r_{h}^{D-2}}{4}  \ ,
\label{eq:(4-3)}
\en
which leads to the relation
\be
dS = \frac{(D-2)\omega_{(D-2)} r_{h}^{D-3}}{4} d r_{h}  \ ,
\label{eq:(4-4)}
\en
and allows to write the first term in the right-hand side of Eq.(\ref{eq:(4-1)}) as
\be
\frac{\partial M}{\partial r_{h}}\Big \vert_{Q} d r_{h} =  \frac{\partial M}{\partial S}\Big \vert_{Q} dS  \ .
\label{eq:(4-5)}
\en

On the other hand, the \textit{surface gravity}, for the spherically symmetric solutions considered here, is defined as \cite{Wald}
\be
k = \frac{1}{2} \frac{\partial g(r)}{\partial r}\Big \vert_{r=r_{h}} = \frac{(D-3)}{2r_{h}} - \omega_{(D-2)} r_{h} {T_0}^{0}  \ .
\label{eq:(4-6)}
\en
We see that Eqs.(\ref{eq:(4-2)}) and (\ref{eq:(4-6)}) are related as
\be
k = \frac{1}{r_{h}^{D-3}} \frac{\partial M}{\partial r_{h}}\Big \vert_{Q} = \frac{(D-2)\omega_{(D-2)}}{4} \frac{\partial M}{\partial S}\Big \vert_{Q}  \ ,
\label{eq:(4-7)}
\en
and lead to the new expression of the differential (\ref{eq:(4-1)}) as
\be
dM = \frac{4 k}{(D-2)\omega_{(D-2)}} d S + \frac{\partial M}{\partial Q}\Big \vert_{r_{h}} d Q = T d S + \frac{\partial M}{\partial Q}\Big \vert_{r_{h}} d Q  \ ,
\label{eq:(4-8)}
\en
where the identification
\be
T = \frac{4 k}{(D-2)\omega_{(D-2)}} = \frac{\partial M}{\partial S}\Big \vert_{Q}  \ ,
\label{eq:(4-9)}
\en
defines the temperature which, as usual, is proportional to the surface gravity (\ref{eq:(4-6)}). From Eq.(\ref{eq:(4-9)}), using Eqs.(\ref{eq:(2-7)}) and (\ref{eq:(2-5)quart}), we obtain the expression:
\bea
T &=& \frac{\Upsilon(D)}{r_{h}} - \frac{4r_{h}}{D-2} (2\varphi_X E^{2} - \varphi)  \nonumber \\
&=& \frac{\Upsilon(D)}{r_{h}} - \frac{8QE}{(D-2)r_{h}^{D-3}} + \frac{4r_{h}}{D-2}\varphi  \ ,
\label{eq:(4-9)bis}
\ena
where the constant $\Upsilon(D)$ is defined as
\be
\Upsilon(D) = \frac{2(D-3)}{(D-2)\omega_{(D-2)} }  \ .
\label{eq:(4-9)ter}
\en
Coming back now to Eq.(\ref{eq:(3-31)}), which defines the extreme BHs, it is obvious from Eq.(\ref{eq:(4-7)}) that both surface gravity and temperature vanish for these configurations\footnote{In $D=4$ dimensions there are some exceptions concerning extreme black points with $T>0$ for some A2 models \cite{dr13}, though no such configurations arise for other families in $D=4$ or for any family in $D>4$ dimensions.}.

The derivative in the second term of the right-hand side of Eq.(\ref{eq:(4-1)}) can be written as
\bea
\frac{\partial M}{\partial Q}\Big \vert_{r_{h}} &=& \frac{\partial \varepsilon_{ex}}{\partial Q}\Big \vert_{r_{h}} =  \omega_{(D-2)} \int_{r_{h}}^{\infty} 2 E(x) dx = \nonumber \\
&=& 2 \omega_{(D-2)} A_{0}(r_{h}) \equiv \Phi(r_{h})  \ ,
\label{eq:(4-10)}
\ena
where Eqs.(\ref{eq:(2-5)quart}), (\ref{eq:(2-26)}) and (\ref{eq:(2-5)penta}) have been used. As easily seen from Eq.(\ref{eq:(2-13)}) this ``normalized electrostatic potential" on the horizon, $\Phi(r_{h}),$ obeys the scale law
\be
\Phi(r_{h},Q,D) = Q^{\frac{1}{D-2}} \Phi(R_{h},Q=1,D)  \ ,
\label{eq:(4-10)bis}
\en
where $R_{h} = r_{h}/Q^{\frac{1}{D-2}}$. The integration performed in Eq.(\ref{eq:(4-10)}) requires the condition $A_{0}(\infty) = 0$. This gauge condition cannot be fulfilled for the elementary solutions of the IRD families, for which the electrostatic potential diverges asymptotically. As already mentioned, an immediate consequence is that the thermodynamic laws cannot be established in these cases, at least in the usual way. In terms of $\Phi(r_{h}),$ the differential (\ref{eq:(4-8)}) becomes
\be
dM = T(S,Q,D) dS + \Phi(S,Q,D) dQ  \ ,
\label{eq:(4-11)}
\en
where the dependence on the two state variables $S$ and $Q$ and the dimension $D$, is made explicit. Eq.(\ref{eq:(4-11)}) is the general expression of the \textit{first law of thermodynamics} for nonrotating charged BHs which are asymptotically Schwarzschild solutions of any admissible NED model minimally coupled to gravity in $D$ spacetime dimensions.

\subsection{The thermodynamic variables and the state diagrams of charged asymptotically flat black holes}

A given non rotating and charged BH configuration is fully characterized by two thermodynamic parameters, whose set of values can be taken as the basis for the elaboration of diagrams displaying other thermodynamic variables as functions of the two chosen ones. One can identify in this way the different BH-states associated to a given NED. The most immediate choice for these parameters are the constants of integration: the mass $M$ and the charge $Q$. In order to characterize BHs belonging to the different families through phase diagrams in terms of these variables, let us split the $Q-M$ plane in several regions through two curves: the $r_{h}=0$ curve and the set of extreme BHs. The constant-$r_{h}$ curves in this $Q-M$ plane are defined by Eq.(\ref{eq:(3-30)}). In particular, the curve $r_{h} = 0$ can be defined only in the cases of the families supporting finite-energy ESS solutions (A1 and A2 families). In the case of the UVD families no such curves exist. In the finite-energy cases, the equation of this curve can be obtained by taking into account the scale law (\ref{eq:(2-31)}), and reads
\bea
M(r_{h}=0,Q,D) &=& \varepsilon_{ex}(r_{h}=0,Q,D)  \nonumber \\
&=& Q^{\frac{D-1}{D-2}} \varepsilon(Q=1,D) \ ,
\label{eq:(4-12)}
\ena
where $\varepsilon(Q=1,D)$ is the electrostatic energy of the field of the unit charge and is a universal constant for a given model with finite-energy solutions. Obviously, $M(r_{h}=0,Q=0,D) = 0$. Moreover, the slope of this curve is given by
\bea
\frac{\partial M}{\partial Q}\Big \vert_{r_{h}=0} &=& \Phi(r_{h}=0,Q,D)  \nonumber \\
&=& Q^{\frac{1}{D-2}} \Phi(r_{h}=0,Q=1,D) \geq 0
\label{eq:(4-13)}
\ena
(see Eqs.(\ref{eq:(4-10)}) and (\ref{eq:(4-10)bis})). This slope is positive for any $Q>0$ and vanishes for $Q=0$, while it diverges at large $Q$ as $Q^{1/(D-2)}$, because in this last equation $\Phi(r_{h}=0,Q=1,D)$ is a universal constant for a given model.

\begin{figure}[t]
 \begin{center}
\includegraphics[width=0.60\textwidth]{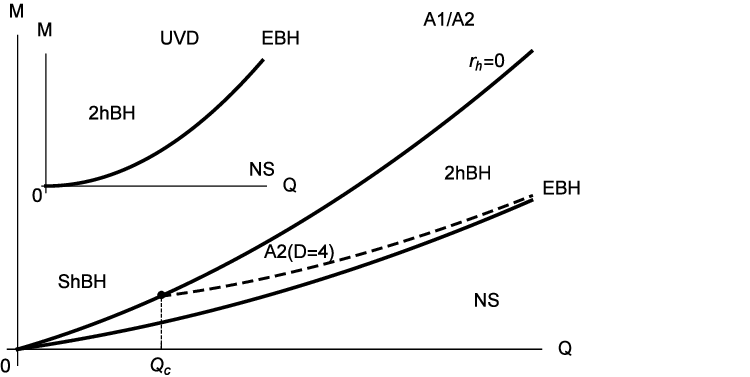}
\caption{Phase diagram in the $Q-M$ plane for typical A1 and A2 models (main frame) and UVD models (small frame) in $D\geq4$. The $r_{h} = 0$ curve in the main frame (corresponding to $M = \varepsilon(Q,D) = Q^{(D-1)/(D-2)} \varepsilon(Q=1,D)$) defines the vanishing inner horizon BH configurations. This curve separates the regions associated to the single horizon (ShBH) and two-horizons (2hBH) black hole configurations. Below the extreme BH curve (EBH) only naked singularity configurations are possible. Both curves meet at the origin, excepted for A2 models in $D=4$, for which the $r_{h} = 0$ curve and the extreme BHs curve (dashed line) meet when the charge takes the critical value $Q_{c} = (16\pi a)^{-1}$ ($a$ being the maximum field strength). In these cases, the piece of the $r_{h} = 0$ curve for $0 < Q \leq Q_{c}$ corresponds to extreme ($Q=Q_{c}$) and  non-extreme ($Q<Q_{c}$) black points, which are absent for all admissible models in $D>4$. The analysis of the diagram in the small frame (UVD cases) is similar, but now the $r_{h} = 0$ curve does not exist and the charged single-horizon BH configurations are absent.}
\label{figure4}
 \end{center}
\end{figure}

The second curve is the set of points $(Q,M_{e})$ corresponding to the extreme BHs. It is obtained by eliminating $r_{h}$ between Eqs.(\ref{eq:(3-30)}) and (\ref{eq:(3-31)}), once the explicit expression of the Lagrangian density is specified. The slope of this curve (which is the $T=0$ isotherm) is given by the derivative
\be
\frac{\partial M}{\partial Q}\Big \vert_{T=0} = \frac{\partial M}{\partial Q}\Big \vert_{r_{h}} + \frac{\partial M}{\partial r_{h}}\Big \vert_{Q} \cdot \frac{\partial r_{h}}{\partial Q}\Big \vert_{T=0} \ .
\label{eq:(4-14)}
\en
The first term in the right-hand side of this formula is the normalized electrostatic potential on the horizon of the extreme BHs ($\Phi(r_{he},Q,D)$). The second term is proportional to the temperature and vanishes for extreme BHs. Thus we have
\be
\frac{\partial M}{\partial Q} \Big\vert_{T=0} = \Phi(r_{he},Q,D) \geq 0 \ .
\label{eq:(4-15)}
\en
Moreover, both curves are tangent to the $Q-$axis on $Q=M=0$. Thus, the main diagram of Fig.\ref{figure4} displays the qualitative behaviour obtained for the A1 and A2 cases in $D>4$ dimensions. These curves in the $Q-M$ plane separate all the different possible phases of the BH states associated to the finite-energy ESS solutions of these families. For models with energy-divergent ESS solutions of the UVD families, the $r_{h}=0$ curves are not defined and the profile of the phase diagram is depicted in the small frame of Fig. \ref{figure4}, where the horizon structure of BHs and naked singularity solutions associated to the different families is apparent.

Let us consider now the temperature function and generalize some important results obtained for this state variable in $D=4$ dimensions. First of all we shall obtain the scale law for the function $T(r_{h},Q,D)$. From the definitions (\ref{eq:(4-7)}) and (\ref{eq:(4-9)}) and the expression of the derivative of the mass parameter given in Eq.(\ref{eq:(3-31)}) we obtain, after some manipulations, the scale relation
\bea
T(r_{h},Q,D) &=& Q^{\frac{1}{D-2}} T(R_{h},Q=1,D)  \nonumber \\
&+& \frac{\Upsilon(D)}{R_{h}}\dfrac{(1-Q^{\frac{2}{D-2}})}{Q^{\frac{1}{D-2}}} \ ,
\label{eq:(4-16)}
\ena
where $R_{h} = r_{h}/Q^{\frac{1}{D-2}}$.

\begin{figure}[h]
 \begin{center}
\includegraphics[width=0.45\textwidth]{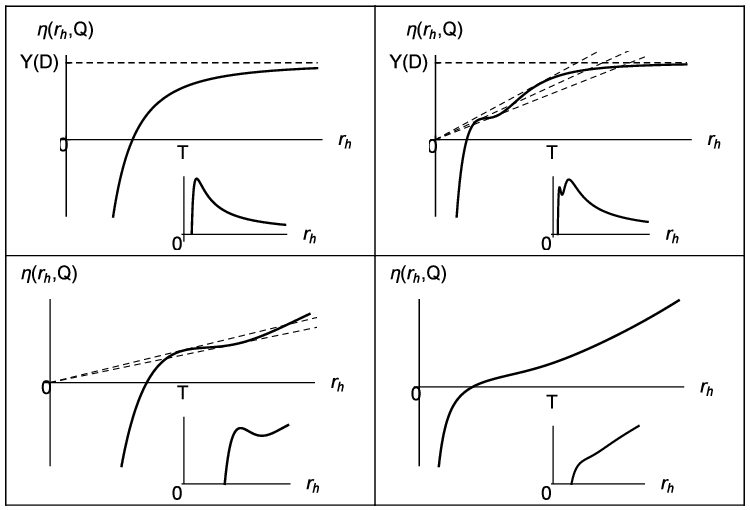}
\caption{Qualitative behaviour of the functions $\eta(r_h)$ obtained from Eq.(\ref{eq:(4-18)}) for asymptotically flat BH solutions associated to two different G-NEDs at constant $Q$ (top figures), and from  Eq.(\ref{eq:(4-23+)}) for two asymptotically AdS black hole solutions corresponding to two sets of values of the parameters $Q$, $D$ and $l$ of a same G-NED (bottom figures). The $\eta(r_h)$ curves are always monotonically increasing and exhibit a vertical asymptote at the origin (excepted for A2 models in $D=4$ dimensions, whose particular features have been extensively analyzed in Ref.\cite{dr13}). In the asymptotically flat cases (top), they exhibit also a horizontal asymptote at the constant value $\eta = \Upsilon(D)$, whereas they diverge parabolically in AdS cases (bottom). The slopes of the dashed straight lines connecting the origin with the points of the $\eta$ curves give the temperatures of the associated BH configurations, which are plotted in the small frames. The tangency points define local or absolute extrema of the $T-r_{h}$ curves and the $\eta = 0$ points correspond to the ($T=0$) extreme BH configurations. The temperature vanishes at large $r_{h}$ for the asymptotically flat BHs and diverges linearly for the asymptotically AdS ones. The models used in obtaining the curves for the asymptotically flat cases are the Born-Infeld one (upper-left) and a UVD-B2 model discussed in Ref.\cite{dr13} (Eq.(73) and Fig.12 of this reference), which exhibits a more rich and complex behaviour of the temperature. For the asymptotically AdS cases the Euler-Heisenberg model has been used for two different sets of parameters, leading also to different qualitative behaviours of the temperature.}
\label{figure5}
 \end{center}
\end{figure}

Concerning the behaviour of the temperature with the horizon radius let us consider the function
\be
\eta(r_{h},Q,D) = r_{h} T(r_{h},Q,D) = \Upsilon(D) - \frac{4 r_{h}^{2}}{D-2} {T_0}^{0} \ .
\label{eq:(4-18)}
\en
It is obvious that, in a $r_{h}-\eta$ diagram, the temperature of the BHs, characterized by their horizon radius at fixed $Q$, equals the slopes of the straight lines that connect the origin and the points of the positive part of the curve $\eta(r_{h},Q,D)$. Using Eq.(\ref{eq:(2-7)}) it can be shown that the last term in the right-hand side of (\ref{eq:(4-18)}) vanishes at large $r_{h}$ for asymptotically normal (Schwarzschild-like) BHs and, consequently, the function $\eta$ exhibits an horizontal asymptote on the value
\be
\eta(r_{h} \rightarrow \infty,Q,D) \rightarrow \Upsilon(D) \ ,
\label{eq:(4-19)}
\en
for all families. In the small-$r_{h}$ region, the curves $\eta(r_{h} \rightarrow 0,Q,D)$, for fixed $Q$, exhibit always a vertical asymptote, due to the divergence of the last term in the right-hand side of Eq.(\ref{eq:(4-18)}) for all families (excepting, as already mentioned, for the A2 models in $D=4$, where $\eta(r_{h}=0,Q,D)$ can be finite). Moreover, the derivation of Eq.(\ref{eq:(4-18)}) yields
\be
\frac{\partial \eta}{\partial r_{h}}\Big \vert_{Q} = - \frac{4}{D-2} \frac{\partial (r_{h}^{2} {T_0}^{0})}{\partial r_{h}}\Big \vert_{Q} \ ,
\label{eq:(4-20)}
\en
which is positive everywhere, as can be seen from Eq.(\ref{eq:(2-11)}) and the first integral (\ref{eq:(2-5)quart}). This way, the curves $\eta(r_{h},Q,D>4)$ for fixed values of $Q$ are monotonically increasing and exhibit the qualitative shapes shown in the upper frames of Fig.\ref{figure5}. As mentioned, the values of the temperature are given by the slopes of the straight lines connecting the origin with the points of the positive part of the $\eta$ curves. They are plotted in the upper small frames of Fig.\ref{figure5} as functions of the horizon radius in two cases. The slopes of the radial lines which are tangent to the $\eta$ curves define local extrema of the temperature and, in particular, the maximum slope tangent determines the absolute maximum of the temperature. At large $r_{h}$, the temperature vanishes asymptotically in all cases. The cut points of the curves with the $\eta=0$ axis correspond to (zero temperature) extreme BHs.

It would be now straightforward to go deeper into the study of the thermodynamic behaviours and properties of the different families of these asymptotically Schwarzschild BHs in $D>4$ dimensions (specific heats, phase transitions, etc) following similar methods as those developed in Ref. \cite{dr13} for $D=4$. Such studies have been carried out for many particular models in the literature, see e.g. \cite{NEDthermodynamics1,
NEDthermodynamics2,NEDthermodynamics3,NEDthermodynamics4,NEDthermodynamics5}. Nevertheless, pursuing our general analysis, we shall henceforth limit our considerations to the study of the relations and the scale laws between the state variables, from which new general and interesting results will arise. As we shall see, this strategy will lead to general methods capturing most of the relevant thermodynamic information of the particular NED cases.

\subsection{Thermodynamics of the asymptotically AdS black holes in $D$ dimensions}

Let us come back to Eq.(\ref{eq:(3-35)}), which gives the $M_{AdS}-r_{h}-Q-l$ relation for asymptotically AdS black holes. If we follow the same steps as in the asymptotically flat cases we must write the differential (\ref{eq:(4-1)}) in terms of the proper variables defining the state functions\footnote{It should be stressed that the consistence of the thermodynamic analysis for asymptotically AdS black holes requires, as in the asymptotically flat cases, the existence of the external energy function $\varepsilon_{ex}(r,Q,D)$ for the underlying NEDs in flat space (B1, B2 and B3 cases).}. Let us first obtain the expression of the derivative of $M_{AdS}$ with respect to $r_{h}$, which reads
\bea
\frac{\partial M_{AdS}}{\partial r_{h}}\Big \vert_{Q} &=& \frac{(D-1)}{2l^{2}} r_{h}^{D-2} + \frac{(D-3)}{2} r_{h}^{D-4} \nonumber \\
&-&  2\omega_{(D-2)} r_{h}^{D-2} {T_0}^{0}  \ .
\label{eq:(4-21+)}
\ena
Using the same definition (\ref{eq:(4-3)}) for the entropy of the asymptotically flat case, we obtain for the first term of the right-hand side of (\ref{eq:(4-1)}) the same expression (\ref{eq:(4-5)}) (with the replacement $M \rightarrow M_{AdS}$), while the temperature is now given by
\bea
T_{AdS}(r_{h},Q,l^{2},D) &=& \frac{\partial M_{AdS}}{\partial S}\Big \vert_{Q} 
\label{eq:(4-22+)} \\
&=& \frac{2(D-1)r_{h}}{(D-2)\omega_{(D-2)}l^{2}} + T_{AF}(r_{h},Q,D) \nonumber \ .
\ena
The first term in the right-hand side of this formula comes from the cosmological term and grows linearly as $r_{h}$ increases, whereas the second term is the expression (\ref{eq:(4-9)bis}) for the temperature in the asymptotically flat case, which vanishes for large-$r_{h}$ BHs and can be continued to unbounded negative values as $r_{h} \rightarrow 0$. This behaviour is well known in several particular models which have been analyzed in the literature, as the generalized Reissner-Nordstr\"om-AdS solutions of the gravitating Maxwell electrodynamics in $D>4$ dimensions \cite{chamblin99}, or the BH solutions associated to the gravitating Born-Infeld-AdS electrodynamics in $D>4$ dimensions, with positive curvature event horizons \cite{dey04}, or negative or zero curvature event horizons \cite{cai04}.

From Eq.(\ref{eq:(4-22+)}) it can be easily shown that the scale law for the temperature takes the same form (\ref{eq:(4-16)}) as in the asymptotically flat case and does no depend explicitly on the cosmological parameter $l^2$, as could be expected from the NED sector origin of the scale invariance. Moreover, the second term of the right-hand side of (\ref{eq:(4-1)}) has the same form (\ref{eq:(4-10)}) as in the asymptotically flat case, and the expression of the second law is the same as in (\ref{eq:(4-11)}) but now we must use the expression (\ref{eq:(4-22+)}) for the temperature. The scale law for $\Phi$ is given by Eq.(\ref{eq:(4-10)bis}) and is explicitly independent of the cosmological parameter.

The function
\bea
\eta_{AdS}(r_{h},Q,l^{2},D) &=& r_{h} T_{AdS} = \dfrac{2(D-1)}{(D-2)\omega_{(D-2)}} \frac{r_{h}^2}{l^2} \nonumber \\
&+& \eta_{AF}(r_{h},Q,D)  \
\label{eq:(4-23+)}
\ena
($\eta_{AF}$ being the asymptotically flat expression (\ref{eq:(4-18)})), exhibits a vertical asymptote at $r_{h} = 0$, and has a monotonically increasing character everywhere, diverging parabolically at large $r_{h}$. Consequently, it cuts once the horizontal axis defining an unique extreme BH. However, the temperature of the large horizon BHs diverges linearly with $r_{h}$ in these asymptotically AdS cases, as expected from the results found in the literature in several particular examples. The qualitative forms of both these $\eta$ functions and the corresponding temperatures in the asymptotically AdS cases are depicted in the bottom frames of Fig.\ref{figure5}.

\section{Thermodynamic relations and scale laws} \label{sectionV}

In this section we shall obtain several relations between the thermodynamic functions and their derivatives. First of all we shall consider the generalization of the well known Smarr formula of the Reissner-Nordstr\"om BHs in $D=4$ to the elementary BH solutions of G-NEDs in $D\geq4$. Next, using the properties of the representations of the scale group in spaces of thermodynamic variables, we shall obtain the induced relations involving the derivatives of the thermodynamic functions and explore some of their consequences.

\subsection{Generalized Smarr formula for asymptotically flat and asymptotically AdS black holes}

A first generalization of the Smarr formula to elementary BH solutions of any G-NED in $D=4$ was carried out in Ref. \cite{dr13} (see also \cite{Breton2005,Gulin,Balart,Zhang:2016ilt,Clement:2017otx}). There, the deviation of the generalized formula in the general NED cases from the simple Smarr formula of the Reissner-Nordstr\"om case was identified in terms of the binding energies associated to the self-interactions of the electric field (due to the nonlinearities of the general NEDs), which contrasts with the linear character of Maxwell electrodynamics. Similar considerations and conclusions are in order here.

Let us consider Eq.(\ref{eq:(4-10)}), which corresponds to the relation between the electrostatic potential on the horizon and the derivative of the exterior energy (\ref{eq:(2-26)}). Using the scale laws (\ref{eq:(2-31)}) we can compute a general expression for this derivative, which reads
\bea
\Phi(r_{h},Q,D) &=& \frac{\partial \varepsilon_{ex}}{\partial Q}\Big \vert_{r_{h}} = \left(\frac{D-1}{D-2}\right) \frac{\varepsilon_{ex}(r_{h},Q,D)}{Q}  \nonumber \\
&+& \frac{\omega_{(D-2)} r_{h}^{D-1} {T_0}^{0}(r_{h},Q,D)}{(D-2)Q} \ .
\label{eq:(5-1)}
\ena
In terms of the ${T_0}^{0}$ component of the energy-momentum tensor, the temperature can be written as
\be
T_{AF} = \frac{\Upsilon(D)}{r_{h}} - \frac{4r_{h}}{(D-2)} {T_0}^{0}(r_{h},Q,D) \ .
\label{eq:(5-2)}
\en
By eliminating ${T_0}^{0}$ in these equations and taking into account the definition (\ref{eq:(4-3)}) of the entropy and the relation (\ref{eq:(3-30)}) we obtain the expression
\be
M_{AF} = \left(\frac{D-2}{D-1}\right) (T_{AF} S + \Phi Q) + \frac{1}{D-1} \left[ \frac{4S}{\omega_{(D-2)}}\right]^{\frac{D-3}{D-2}} \ ,
\label{eq:(5-3)}
\en
This is a finite relation between the thermodynamic variables involved in the first law and generalizes the Smarr formula of the Reissner-Nordstr\"om BHs in $D=4$ to the elementary BH solutions of any G-NED in $D \geq 4$ spacetime dimensions. Indeed, in the Maxwell case it reduces to the well known Smarr expression of the D-dimensional extension of the Reissner-Nordstr\"om BH. For other particular models studied in the literature, such as the Euler-Heisenberg and Born-Infeld ones or the family of power Maxwell Lagrangian densities ($\varphi(X) = X^{p}, p$ being a positive integer \cite{hassaine09-NED}) the restriction of the general formula (\ref{eq:(5-3)}) naturally yields the correct particular expressions for the respective Smarr formulae.

For asymptotically AdS black holes a generalized Smarr-like formula can be obtained too. Starting now with Eqs.(\ref{eq:(3-33)}) and (\ref{eq:(3-35)}) and following the previous steps we arrive to the expression
\bea
M_{AdS} &=& \left(\frac{D-2}{D-1}\right) (T_{AdS} S + \Phi Q) + \frac{1}{D-1} \left[ \frac{4S}{\omega_{(D-2)}}\right]^{\frac{D-3}{D-2}},
\label{eq:(5-3)bis}
\ena
for the general Smarr-like formula for asymptotically AdS black holes which are ESS solutions of Einstein's field equations with cosmological term and minimally coupled to a general NED with topologically spherical horizons. This expression is formally identical to Eq.(\ref{eq:(5-3)}), but now the dependence of the mass on the cosmological term is implicit in the expression (\ref{eq:(4-22+)}) of the temperature $T_{AdS}$.

\subsection{Scale laws and scale group for asymptotically flat black holes}

Let us first consider the expressions of the scale laws under the form of explicit one-parameter transformations of the different state variables. The electrostatic field and the energy, referred to the horizon, scale as in Eqs.(\ref{eq:(2-14)}) and (\ref{eq:(2-32)}) respectively, when $r=r_{h}$. The scale law for the mass is given in Eq.(\ref{eq:(3-30)ter}). The scale law for the normalized electrostatic potential on the horizon is obtained from Eq.(\ref{eq:(4-10)bis}) and reads
\be
\Phi(\theta r_{h},\theta^{D-2}Q,D) = \theta \Phi(r_{h},Q,D) \ .
\label{eq:(5-5)}
\en
For the temperature, the one-parameter scale transformations result from Eq.(\ref{eq:(4-16)}), and read
\be
T(\theta r_{h},\theta^{D-2}Q,D) = \theta T(r_{h},Q,D) + \frac{(1-\theta^{2})}{\theta} \frac{\Upsilon(D)}{r_{h}} \ .
\label{eq:(5-6)}
\en
Obviously, the entropy scales as
\be
S(\theta r_{h},D) = \theta^{D-2} S(r_{h},D) \ .
\label{eq:(5-7)}
\en
The form of the scale laws defined by these equations can be interpreted as different representations of the group of the transformations $\Gamma(\theta)$ in the spaces of the corresponding state functions and independent variables (in present cases $Q$ and $r_{h}$). It is, indeed, straightforward to verify that the group laws (\ref{eq:(2-15)V}) are satisfied by these representations. Similar scale laws involving other thermodynamic functions and variables can be easily obtained. By using Eqs.(\ref{eq:(4-3)}) and (\ref{eq:(5-7)}) we can write the corresponding representations in terms of the independent variables $Q$ and $S$. In particular, Eq.(\ref{eq:(3-30)ter}) becomes
\bea
M(\theta^{D-2}S,\theta^{D-2}Q,D) &=& \theta^{D-1} M(S,Q,D) \label{eq:(5-4)}  \\
&+& \frac{\theta^{D-3}(1-\theta^{2})}{2} \left(\dfrac{4S}{\omega_{(D-2)}}\right)^{\frac{D-3}{D-2}}\nonumber   \ ,
\ena
where the functional dependence in the entropy of the mass $M(S,Q,D)$ is obtained from $M(r_{h}(S),Q,D)$ with $r_{h}(S)$ given by Eq.(\ref{eq:(4-3)}). With the same notation, the scale transformation for the potential $\Phi(S,Q)$ takes the form
\be
\Phi(\theta^{D-2}S,\theta^{D-2}Q,D) = \theta \Phi(S,Q,D) \
\label{eq:(5-4)bis}
\en
and for the temperature $T(S,Q)$
\bea
T(\theta^{D-2}S,\theta^{D-2}Q,D) &=& \theta T(S,Q,D) \label{eq:(5-4)ter} \\
&+& \frac{(1-\theta^{2})}{\theta} \Upsilon(D) \left(\dfrac{4S}{\omega_{(D-2)}}\right)^{\frac{-1}{D-2}} \nonumber \ .
\ena
The infinitesimal generators of this scale group on each representation are obtained by deriving the explicit form of the transformations with respect to the parameter $\theta$ on the identity ($\theta=1$). In the case of Eq.(\ref{eq:(5-4)}) we obtain the following expression for the scale group equation
\bea
Q \dfrac{\partial M}{\partial Q}\Big\vert_{S} &+& S \dfrac{\partial M}{\partial S}\Big\vert_{Q} - \dfrac{D-1}{D-2} M \label{eq:(5-8)} \\
&+& \dfrac{1}{D-2}  \left(\dfrac{4S}{\omega_{(D-2)}}\right)^{\frac{D-3}{D-2}} = 0 \ . \nonumber
\ena
We note that, by replacing the definitions (\ref{eq:(4-9)}) and (\ref{eq:(4-10)}) in this equation, we recover the general Smarr formula (\ref{eq:(5-3)}). This is an alternative way in obtaining this formula, which appears now as a ``renormalization group"-like equation whose origin lies in the internal symmetry $\Gamma(\theta)$ fulfilled by any NED. As we shall see at once, the same procedure can be used in obtaining the corresponding formula (\ref{eq:(5-3)bis}) for the asymptotically AdS cases. This same expression will remain valid even for extensions of the BH thermodynamics which include the cosmological constant as a state function (see subsection D below).

For the normalized electrostatic potential on the horizon $\Phi_{h}(S,Q)$ the corresponding scale group equation results from Eq.(\ref{eq:(5-4)bis}) and reads
\be
Q \dfrac{\partial \Phi_{h}}{\partial Q}\Big\vert_{S} + S \dfrac{\partial \Phi_{h}}{\partial S}\Big\vert_{Q} - \dfrac{\Phi_{h}}{D-2} = 0 \ .
\label{eq:(5-9)}
\en
and for the temperature $T(S,Q)$ we obtain from (\ref{eq:(5-4)ter})
\be
Q \dfrac{\partial T}{\partial Q}\Big\vert_{S} + S \dfrac{\partial T}{\partial S}\Big\vert_{Q} - \dfrac{T}{D-2} + \dfrac{2\Upsilon(D)}{D-2} \ . \left[\dfrac{\omega_{(D-2)}}{4S}\right]^{\frac{1}{D-2}} = 0.
\label{eq:(5-10)}
\en
These equations deserve some analysis. First of all, they are independent of the Lagrangian functions defining the particular NEDs and, in this sense, they are ``universal" laws of the BH thermodynamics in this context. They do not explicitly involve the gravitational sector of the models but come instead from the scale invariance (\ref{eq:(2-15)}) of the NED sector. Moreover, they are linear, first-order, partial differential equations relating derivatives of the thermodynamic functions ($M(S,Q), \Phi_{h}(S,Q)$ and $T(S,Q)$ in the present examples) and have the generic form
\be
Q \dfrac{\partial Z}{\partial Q}\Big\vert_{S} + S \dfrac{\partial Z}{\partial S}\Big\vert_{Q} + \alpha Z + \beta S^{\gamma} = 0 \ ,
\label{eq:(5-11)}
\en
where $Z$ must be identified with any of these functions and the constants $\alpha, \beta,$ and $\gamma$ are immediately identified from Eqs.(\ref{eq:(5-8)})-(\ref{eq:(5-10)}) in each case.

In solving Eq.(\ref{eq:(5-11)}) we can obtain the associated beam of characteristics in the $Q-S-Z$ space, which are solutions of the differential system
\be
\frac{dQ}{Q} = \frac{dS}{S} = \dfrac{-dZ}{\alpha Z+\beta S^{\gamma}} \ .
\label{eq:(5-12)}
\en
The general solution of this system in parametric form, in terms of a parameter $\tau > 0$, is
\be
Q = Q_{0} \tau\hspace{.1cm} ; \hspace{.1cm} S = S_{0} \tau \hspace{.1cm}; \hspace{.1cm} Z = Z_{0} \tau^{-\alpha} - \dfrac{\beta S_{0}^{\gamma}}{\alpha+\gamma} \left(\tau^{\gamma} - \tau^{-\alpha}\right) \ ,
\label{eq:(5-13)}
\en
where the integration constants $Q_{0}, S_{0}$ and $Z_{0}$ are the coordinates of points in the $Q-S-Z$ space defining the particular characteristic it belongs to (for $\tau = 1$). These curves lie on planes of the beam $S = \frac{S_{0}}{Q_{0}}Q$, which can be characterized by the angle $\vartheta$ they form with the $Q-Z$ plane. On each one of these planes we can introduce, besides the coordinate $Z$, the coordinate $\xi$ defined by
\be
\xi = \sqrt{Q^{2} + S^{2}} = \tau \sqrt{Q_{0}^{2} + S_{0}^{2}} \ ,
\label{eq:(5-14)}
\en
($\xi, \vartheta$ and $Z$ are cylindrical coordinates in the $Q-S-Z$ space) and the equations of the characteristics on these constant-$\vartheta$ planes read
\be
Z = Z_{0} \left( \frac{\xi}{\xi_{0}}\right) ^{-\alpha} - \dfrac{\beta (\xi_{0}\sin(\vartheta))^{\gamma}}{\alpha+\gamma} \left[ \left( \frac{\xi}{\xi_{0}}\right) ^{\gamma} - \left( \frac{\xi_{0}}{\xi}\right) ^{\alpha}\right] \ ,
\label{eq:(5-15)}
\en
for $0 < \vartheta \leq \pi/2$, and by
\be
Z = Z_{0} \left( \frac{\xi}{\xi_{0}}\right)^{-\alpha} = Z_{0} \left( \frac{Q}{Q_{0}}\right)^{-\alpha} \ ,
\label{eq:(5-16)}
\en
for $\vartheta = 0$. The limit of Eq.(\ref{eq:(5-15)}) as $\vartheta \rightarrow 0$ (which implies $S \rightarrow 0$) is singular and the exact expression (\ref{eq:(5-16)}) can never be reached by any sequence of decreasing-entropy BHs.

The BHs associated to a given admissible G-NED are characterized by the values of two thermodynamic functions (e.g. $Q,S$), in terms of which other thermodynamic functions (e.g. $Z \equiv M,T,\Phi, \ldots$) can be determined through an equation of state (EOS): $Z=Z(Q,S)$, which must be a solution of the ``universal" equation (\ref{eq:(5-11)}) and defines a surface in the $Q-S-Z$ space, whose points (in the physically meaningful regions) correspond to the BH solutions of the model. Such surfaces are generated by the beam of characteristics (\ref{eq:(5-13)}). In order to determine the particular surface associated with a given model we can look for a non-characteristic line belonging to this surface. A simple strategy is to work with the variables $Q-S-T$. In this case, once the explicit form of the Lagrangian density is specified, the set of extreme BHs ($T=0$) defines a curve in the $Q-S$ plane through Eq.(\ref{eq:(3-32)bis}). In this way the EOS $T=T(Q,S)$ can be explicitly determined for each model. Using these variables, the values of the parameters $\alpha, \beta$ and $\gamma$ are
\be
\alpha = \gamma = -\frac{1}{D-2}\hspace{.1cm}; \hspace{.1cm} \beta = \dfrac{2\Upsilon(D)}{D-2} \left(\dfrac{\omega_{(D-2)}}{4}\right)^{\frac{1}{D-2}} \ ,
\label{eq:(5-17)}
\en
(see Eq.(\ref{eq:(5-10)})) and the equations of the characteristics become
\bea
T &=& T_{0} \left( \frac{\xi}{\xi_{0}}\right) ^{\frac{1}{D-2}} - \Upsilon(D)\Big(\dfrac{\omega_{(D-2)}}{4\xi_{0}\sin(\vartheta)}\Big)^{\frac{1}{D-2}} \nonumber \\ &\times& \Big[\Big(\frac{\xi}{\xi_{0}}\Big)^{\frac{1}{D-2}} - \Big(\frac{\xi_{0}}{\xi}\Big)^{\frac{1}{D-2}}\Big] \ ,
\label{eq:(5-18)}
\ena
for $0 < \vartheta \leq \pi/2$ and
\be
T = T_{0} \left( \frac{\xi}{\xi_{0}}\right)^{\frac{1}{D-2}} = T_{0}\left( \frac{Q}{Q_{0}}\right)^{\frac{1}{D-2}} \ ,
\label{eq:(5-19)}
\en
for $\vartheta = 0$, this last equation corresponding to the exact $S=0$ state which will never be reached.

\begin{figure}[t]
 \begin{center}
\includegraphics[height=6.0cm,width=8.5cm]{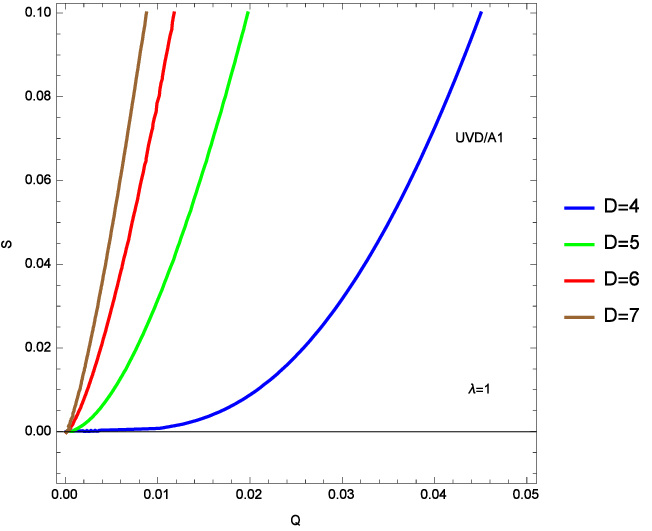}
\includegraphics[height=6.0cm,width=8.5cm]{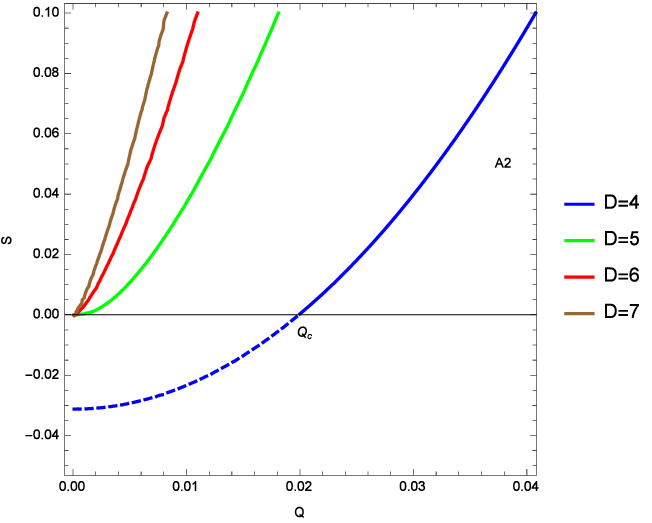}
\caption{Qualitative behaviours of the curves of extreme BHs in the $Q-S$ plane for the three families of central-field behaviours. The UVD and A1 cases (upper frame) and the A2 case (lower frame) are represented for several values of the spacetime dimension $D$. The curves are positive definite everywhere, excepting for the A2 family in $D=4$, where extreme and non-extreme black points are present (in ($Q=Q_{c}, S=0$) and ($Q<Q_{c}, S=0$), respectively). The model used in calculating the curves of the upper frame is the Euler-Heisenberg one, belonging to the A1 family in $D=4$ and to the UVD family in $D>4$. The lower frame is obtained from the Born-Infeld model, as representative of the A2 family in any dimension.}
\label{figure6}
 \end{center}
\end{figure}

Let us come back now to the determination of the curves defining the extreme BHs in the $Q-S$ plane. By using the relation
\be
R_{h} \equiv \frac{r_{h}}{Q^{\frac{1}{D-2}}} = \left( \frac{4S}{\omega_{D-2}Q}\right)^{\frac{1}{D-2}} \ ,
\label{eq:(5-20)}
\en
then Eq.(\ref{eq:(3-32)bis}) can be rewritten in terms of the entropy as
\be
\left( \frac{4S}{\omega_{(D-2)}}\right)^{\frac{D-4}{D-2}} = \frac{4}{D-3} (\omega_{(D-2)}QE - 2S\varphi) \ .
\label{eq:(5-21)}
\en
Once the Lagrangian density $\varphi(X)$ is specified this equation defines the curve
\be
f(Q,S_{ext}) = 0 \ ,
\label{eq:(5-22)}
\en
in implicit form. The analysis of Eq.(\ref{eq:(5-21)}) and its derivative (by using the central and asymptotic behaviours of $E(r)$ and $\varphi(X)$ through Eqs.(\ref{eq:(2-21)})-(\ref{eq:(2-23)}) and the consequences of the admissibility conditions discussed in Section \ref{sectionII}) shows that the function $S_{ext}(Q)$ vanishes when $Q \rightarrow 0$, and is positive and monotonically increasing for any value of $Q > 0$ for all admissible models in $D>4$ spacetime dimensions, as well as for A1 and UVD models in $D=4$. For models of the family A2 in $D=4$ this function vanishes for the critical value of the charge $Q=Q_{c}=1/(16\pi a)$, corresponding to the extreme black points. In this last case, $S_{ext}(Q)$ becomes negative for $Q < Q_{c}$ (there are not extreme BHs in this range of charges). In Fig.\ref{figure6} we have depicted the curves of the extreme BHs in the $Q-S$ plane for the different families of central-field behaviours and several values of the spacetime dimension $D$. The special behaviour of the A2 family in $D=4$ dimensions is apparent on the lower frame.

It is now easy to outline the analytical procedure allowing to build the EOS of the full set of BH solutions associated with a given admissible G-NED, once the expression of the Lagrangian density $\varphi(X)$ is known. The EOS function $T = T(Q,S)$ defines a surface in the $Q-S-T$ space and is obtained by eliminating $Q_{0}$, $S_{0}$ (with $T_{0} = 0$) between Eqs.(\ref{eq:(5-18)}) and (\ref{eq:(5-22)}) (written in terms of $Q_{0}$ and $S_{0}$). Since all the characteristics in the constant-$\vartheta$ planes are asymptotic to the temperature-axis (see Eq.(\ref{eq:(5-18)}): $T \rightarrow \infty$ as $\xi \rightarrow 0$), it is obvious that the points of these EOS surfaces exhaust the full set of BHs associated to a given model. Moreover, the analysis of the thermodynamic behaviour of the BH solutions of a given model can be split into two parts. On the one hand, those properties of the EOS surface coming from the structure of the beam of characteristics, which is independent of the particular NED chosen. On the other hand, the properties induced by the structure of the extreme BHs line, which differ for each model.

\begin{figure}[t]
 \begin{center}
\includegraphics[height=6.5cm,width=8.5cm]{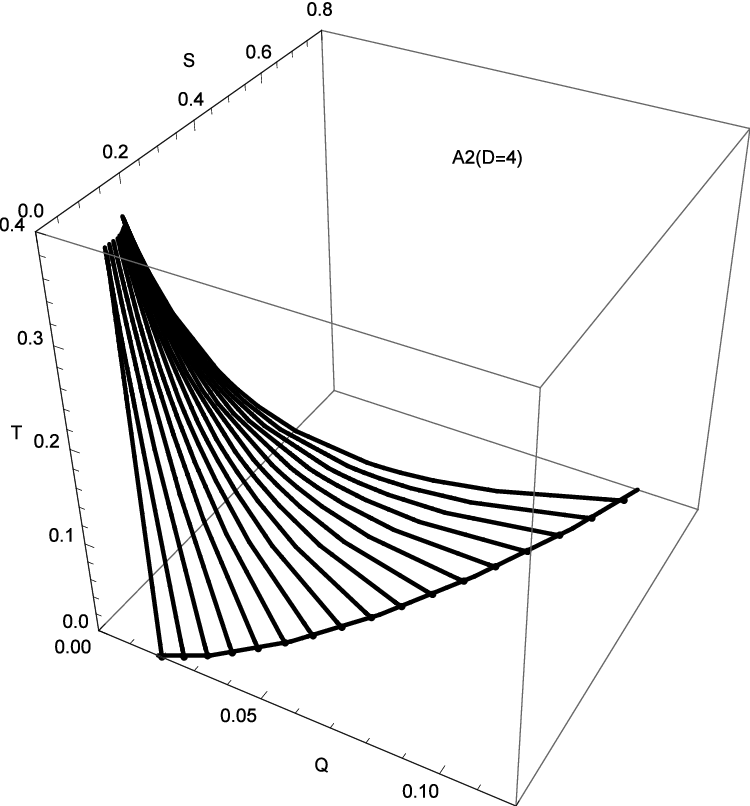}
\includegraphics[height=6.5cm,width=8.5cm]{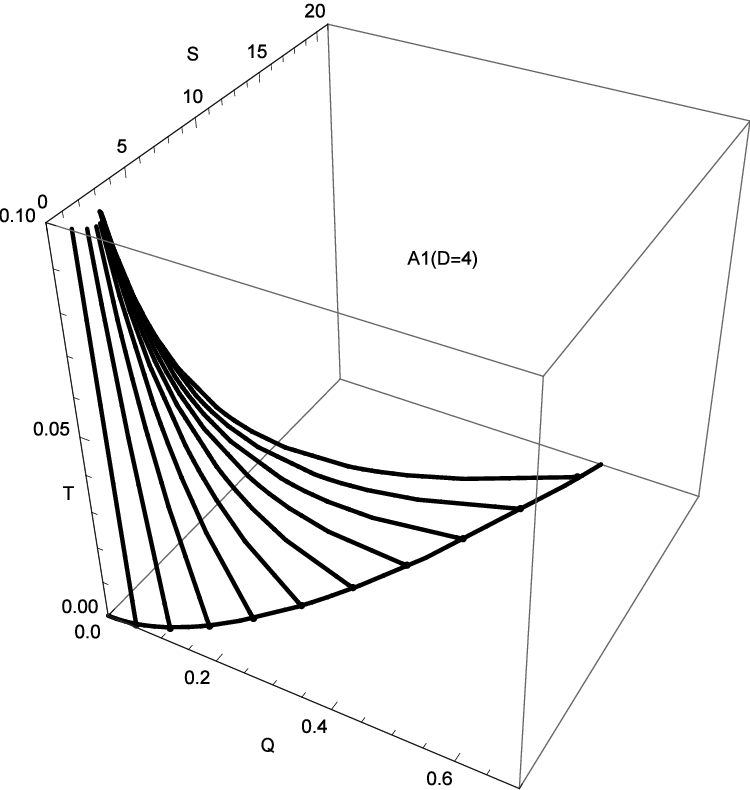}
\caption{Qualitative shapes of the EOS surfaces built by the method of the characteristics associated to the sets of BH solutions of different families of G-NED models. The top frame displays the typical behaviour of the families A2 in $D=4$ and is obtained here using the Born-Infeld model. The bottom frame shows the qualitative behaviour for the other families in any dimension (here the example used is the Euler-Heisenberg model in $D=4$).}
\label{figure7}
 \end{center}
\end{figure}

Fig.\ref{figure7} shows the qualitative shape of the EOS surfaces generated by the characteristic beam from the set of extreme BHs for two particular cases. The upper frame corresponds to the special cases of A2 models in $D=4$, whereas the lower frame displays the typical behaviour for non-A2 models in $D=4$ and for all models in $D>4$. Obviously, the quantitative behaviour depends of the particular model through the explicit form of their associated extreme BHs line.

\subsection{Scale laws and scale group for asymptotically AdS black holes}

Let us consider the representations of the scale group for the different thermodynamic variables of the asymptotically AdS black hole solutions. As can be easily verified, the scale laws for the state variables $M_{AdS}, \Phi$ and $T_{AdS}$, as functions of $S$, $Q$ and $l$ in these representations, are explicitly independent of the cosmological term and coincide formally with those established in Eqs.(\ref{eq:(5-4)}), (\ref{eq:(5-4)bis}) and (\ref{eq:(5-4)ter}) for the asymptotically flat cases (with the replacements $M \rightarrow M_{AdS}$ and $T \rightarrow T_{AdS}$). As a consequence, the $\Gamma(\theta)$ parametric group transformations, as well as the generating equations (\ref{eq:(5-8)})-(\ref{eq:(5-10)}) (with the same replacements) take the same form in both cases, confirming, in particular, the formal identity between the corresponding generalized Smarr formulae. The beams of characteristics associated to these equations are also identical in the corresponding three dimensional spaces of the involved thermodynamic variables. Moreover, the beam of characteristics in the $Q-S-T_{AdS}$ space is determined by Eqs.(\ref{eq:(5-18)}) and (\ref{eq:(5-19)}) (with the replacement $T \rightarrow T_{AdS}$) which coincide formally in both asymptotically flat and AdS cases.

\begin{figure}[t]
 \begin{center}
\includegraphics[height=5.2cm,width=8.5cm]{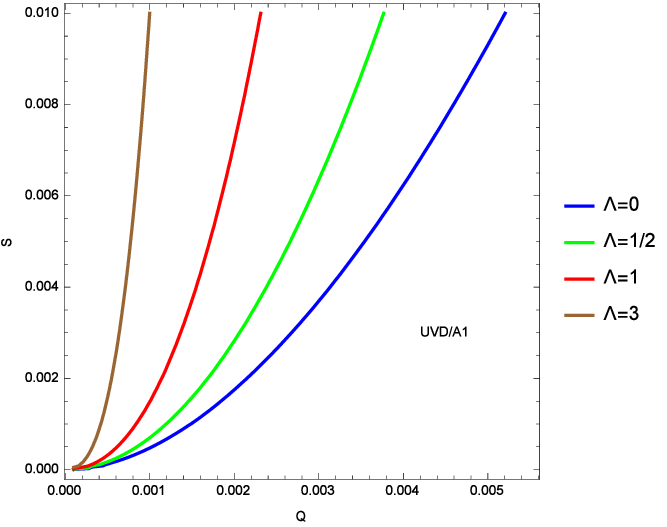}
\includegraphics[height=5.2cm,width=8.5cm]{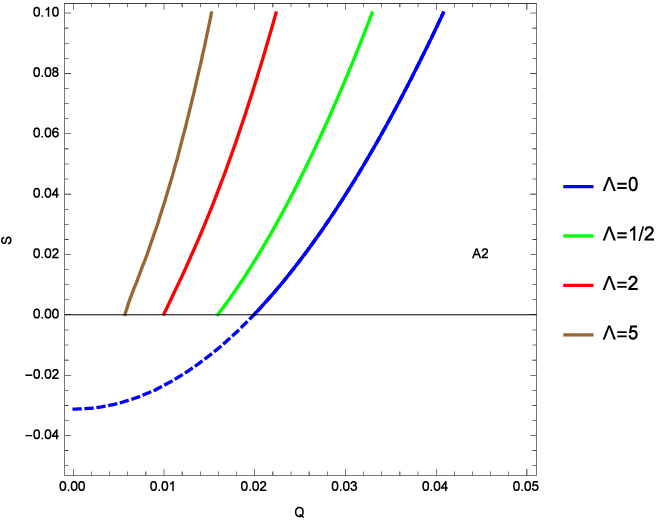}
\caption{Quantitative behaviour of  asymptotically AdS extreme BHs for several values of the cosmological constant in the $Q-S$ plane. The upper frame corresponds to the gravitating Euler-Heisenberg model in $D=5$, as a representative of the UVD models in higher dimensions. The lower frame comes from the gravitating Born-Infeld model, as a representative of the special case of A2 models in $D=4$. In these last cases, the continuation of the curves in the unphysical region ($S<0$) has been removed, excepting for the asymptotically flat curve $\Lambda = 0$ (dashed piece).}
\label{figure8}
 \end{center}
\end{figure}

Thus, following the same procedure used for the asymptotically flat cases, in obtaining the EOS of the set of asymptotically AdS black holes associated to a given NED model in the $Q,S,T_{AdS}$ space, the knowledge of the curve of asymptotically AdS extreme BHs is required. The equation of this curve in the $Q-S$ plane can be obtained from Eqs.(\ref{eq:(3-40)}) and (\ref{eq:(5-20)}) and reads
\bea
\Bigg( \frac{4S}{\omega_{(D-2)}}\Bigg)^{\frac{D-4}{D-2}} &=& \frac{4}{D-3} \Bigg[ \omega_{(D-2)}QE \nonumber \\
&-&
2S\Bigg(\varphi + \frac{(D-1)}{2\omega_{(D-2)} l^{2}} \Bigg)\Bigg] \ .
\label{eq:(5-23)}
\ena
This expression must be compared with Eq.(\ref{eq:(5-21)}). We see that the effect of the cosmological term is to add a constant to the Lagrangian density function which, at first sight, should not introduce important qualitative modifications in the form of the extreme BH curves. Figure \ref{figure8} exhibits the typical behaviour of these curves corresponding to the asymptotically flat and AdS black holes solutions associated to two given models and dimensions (Euler-Heisenberg in $D=5$, as a UVD model, and Born-Infeld in $D=4$, as an A2 model) for several values of the cosmological constant. We see that the quantitative effect of the cosmological term is to increase the value of the entropy (or the horizon radius) of the extreme BHs for fixed values of the charge, but no other qualitative new features seems to arise from the presence of this term in the physically meaningful region ($S>0$).

Because the beam of characteristics is similar in the asymptotically flat and AdS cases, the form of the EOS surfaces for the BH solutions of a given model, with and without a cosmological term, is affected only by the differences between the extreme BH curves. In general, they should be qualitatively similar in both cases, exhibiting shapes as in Fig.\ref{figure7}. Nevertheless, important qualitative differences can not be excluded for some particular models.

\subsection{Extensions of black hole thermodynamics}

Extensions of BH thermodynamics have been proposed and analyzed in the literature in the last years. They are mainly motivated by their eventual usefulness in the context of the AdS/CFT correspondence. The first one concerns the extension of the phase space of the asymptotically AdS BH degrees of freedom by including the cosmological constant as a new thermodynamic variable \cite{Teitelboim1985,kastor2009,mann2012bis,mann2017}. The second one concerns the particular case of the gravitating Born-Infeld model and the extension consists in the assumption that the maximum field strength (the parameter $a = 1/\mu$ in Eq.(\ref{eq:(2-38)})) becomes a new thermodynamic variable \cite{mann2012}.  Although the systematic analysis of these extensions goes beyond the purposes of this work, let us consider some consequences of our methods for the first of these problems.

The expressions of the generalized Smarr formulae (\ref{eq:(5-3)}) and (\ref{eq:(5-3)bis}) for the NED-based, elementary, asymptotically flat or AdS BHs, have been deduced directly from the thermodynamic formulae for the mass, the temperature and the potential $\Phi(r_{h})$, in every case. As already mentioned, the law (\ref{eq:(5-3)}) in the asymptotically flat cases can be deduced alternatively from the scale invariance under the group $\Gamma(\theta)$, whose representation in the $Q-S-M_{AF}$ space is given by Eq.(\ref{eq:(5-4)}) (with the replacement $M \rightarrow M_{AF}$). A similar procedure can be used for an alternative deduction of (\ref{eq:(5-3)bis}) in the asymptotically AdS cases. Indeed, the representation of the $\Gamma(\theta)$ group in the $Q-S-M_{AdS}$ space is obtained from Eqs.(\ref{eq:(3-35)}) and (\ref{eq:(4-3)}) through the group transformations of the independent variables: $S \rightarrow \theta^{D-2}S$, $Q \rightarrow \theta^{D-2}Q$ whereas $l$ remains constant. This way we obtain
\bea
&&M_{AdS}(\theta^{D-2}S,\theta^{D-2}Q,l,D) = \theta^{D-1} M_{AdS}(S,Q,l,D) \nonumber  \\
&+& \frac{\theta^{D-3}(1-\theta^{2})}{2} \left(\dfrac{4S}{\omega_{(D-2)}}\right)^{\frac{D-3}{D-2}},
\label{eq:(5-24)}
\ena
where the explicit dependence on the cosmological parameter $l$ disappears. This expression is formally identical to the asymptotically flat expression (\ref{eq:(5-4)}). The derivative of this equation with respect to $\theta$ in $\theta=1$, together with the definitions (\ref{eq:(4-22+)}) of $T_{AdS}$ and (\ref{eq:(4-10)}) of $\Phi$ (with the replacement $M \rightarrow M_{AdS})$), leads directly to the Smarr formula in Eq.(\ref{eq:(5-3)bis}).

For a variable cosmological constant the first law takes the form
\bea
dM_{AdS} &=& T_{AdS}(S,Q,l,D) dS + \Phi(S,Q,l,D) dQ  \nonumber \\
&+& \frac{\partial M_{AdS}}{\partial \Lambda}d\Lambda \label{eq:(5-25)} \\
&=& T_{AdS} dS + \Phi dQ - \frac{1}{2}\frac{\partial M_{AdS}}{\partial l} dl \nonumber  \ .
\ena
In obtaining a generalized Smarr formula in this case, we can extend the $\Gamma(\theta)$ group by including transformations of the variable $l$ in such a way that the independent variables transform as $S \rightarrow \theta^{D-2}S$, $Q \rightarrow \theta^{D-2}Q$ and $l \rightarrow \theta l$. Now the law of transformation for the mass becomes:
\bea
&&M_{AdS}(\theta^{D-2} S,\theta^{D-2} Q,\theta l,D) = \theta^{D-1} M_{AdS}(S,Q,l,D) \nonumber  \\
&+& \frac{\theta^{D-3}(1-\theta^{2})}{2} r_{h}(S)^{D-3}\left(1 + \dfrac{r_{h}(S)^{2}}{l^{2}}\right),
\label{eq:(5-26)}
\ena
where we have introduced the notation
\be
r_{h}(S) = \left( \dfrac{4S}{\omega_{(D-2)}}\right)^{\frac{1}{D-2}} \ ,
\label{eq:(5-26)bis}
\en
by simplicity. By deriving this equation with respect to $\theta$ in $\theta=1$, the explicit dependence in $l$ disappears and using the definitions (\ref{eq:(4-22+)}) and (\ref{eq:(4-10)}) of $T_{AdS}$ and $\Phi$, the final expression coincides with the generalized Smarr formula (\ref{eq:(5-3)bis}) of the constant $l$ case.

Moreover, we can obtain the same results by following the usual scaling argument associated to \textit{dimensional} homogeneity\footnote{But now the Euler theorem cannot be directly used, because the \textit{functional} homogeneity of $M_{AdS}(S,Q,l,D)$ does not hold, in general.}. Indeed, in this case the independent variables have the dimensions $[S]\sim L^{D-2}, [Q] \sim L^{D-3}$ and $[l] \sim [r_{h}] \sim L$ and, in a length dilatation of amplitude $\theta$, scale as $S \rightarrow \theta^{D-2} S$, $Q \rightarrow \theta^{D-3} Q$, $r_{h} \rightarrow \theta r_{h}$ and $l \rightarrow \theta l$, leading to
\bea
&&M_{AdS}(\theta^{D-2}S,\theta^{D-3}Q,\theta,D) = \theta^{D-1} M_{AdS}(S,\frac{Q}{\theta},l,D) \nonumber  \\
&+& \frac{\theta^{D-3}(1-\theta^{2})}{2} r_{h}(S)^{D-3}\left(1 + \dfrac{r_{h}(S)^{2}}{l^{2}}\right)
\label{eq:(5-27)}
\ena
By deriving this expression with respect to $\theta$ in $\theta = 1$ we recover the expression of the generalized Smarr formula (\ref{eq:(5-3)bis}).
In the asymptotically flat cases one can confirm the validity of Eq.(\ref{eq:(5-3)}) with similar scaling arguments, both using the $\Gamma(\theta)$ group invariance (as already done in Eqs.(\ref{eq:(5-4)}) and (\ref{eq:(5-8)})) or the dimensional argument.

We conclude that Eqs.(\ref{eq:(5-3)}) and (\ref{eq:(5-3)bis}) are robust universal relations valid for all the gravitating NEDs and involving the thermodynamic functions entering in the first law (even in the case in which the cosmological length is treated as a thermodynamic function). Obviously, these relations reduce to different expressions of the Smarr formulae for particular models, once the forms of their Lagrangian densities are specified and the external energy functions $\epsilon_{ex}(S,Q,D)$ (and other particular relations between the thermodynamic functions) are explicitly determined in every case. As already mentioned, one can verify that the particular Smarr formulae found in the literature for several particular models can be recovered in this way from the general expressions (\ref{eq:(5-3)}) and (\ref{eq:(5-3)bis}).

The second possible extension of the thermodynamics concerns the treatment of coupling constants involved in the Lagrangian densities of NEDs as new thermodynamic variables. As already mentioned, this has been performed for the particular case of the Born-Infeld model \cite{mann2012}. In the general case of admissible models characterized by Lagrangians of the form $\varphi(X,\mu_{i})$, where $\mu_{i}$ are a finite sequence of parameters, the first law for elementary BH solutions could be, in principle, generalized to the form
\begin{eqnarray}
dM &=& T_{AdS}(S,Q,l,\mu_{i},D) dS + \Phi(S,Q,l,\mu_{i},D) dQ \nonumber  \\
&+& \Sigma_{i}\frac{\partial M}{\partial \mu_{i}} d\mu_{i}.
\label{eq:(5-28)}
\end{eqnarray}
At first sight it seems difficult that general laws for this extended problem exist. Nevertheless, we are exploring this question beyond the particular case of the gravitating BI electrodynamics and it seems that, under some suitable conditions, one can obtain families of models exhibiting interesting extended thermodynamic properties under this scaling which deserve to be analyzed. But this will be matter of future work.

\section{Conclusion and perspectives} \label{sectionVI}

In this work we have considered the structural and thermodynamic properties of both asymptotically flat and Anti-de-Sitter elementary black hole solutions resulting from the minimal coupling of general nonlinear electrodynamic models, in $D \geq 4$ spacetime dimensions, to the gravitational field (including or not a cosmological term). These models were constrained by several requirements endorsing their physical consistency. Next, they were classified in several families according to the behaviour of their Lagrangian densities in vacuum and at the boundary of their domain of definition (or, equivalently, by the asymptotic and central field behaviours of their elementary solutions). This classification exhausts the set of physically meaningful NEDs in $D \geq 4$ spacetime dimensions.

The heart of the methods developed in this paper lies on the fact that, when coupled to gravitation, such constraints and classifications allow us for a full characterization of the structural and thermodynamic properties of the BH solutions corresponding to the different families, without providing the explicit expression of the Lagrangian density function defining every particular NED model. Once this function is specified, the methods and general formulae provided here allow one to obtain the detailed behaviours on each particular case.

For the structural properties we have integrated the field equations assuming topologically spherical horizons only. Then we have split the problem into the asymptotically flat and AdS cases. For the former, our analysis reveals that the only BH configurations allowed in this setting have either one event horizon, two horizons, or a single degenerate horizon (extreme BHs). The single (non-degenerate) horizon BHs arise only for the models for which the total electrostatic energy of the configurations ($\epsilon(Q,D)$) is finite (soliton supporting models) and when the BH mass exceeds this value ($M>\epsilon(Q,D)$). Otherwise, the BHs exhibit always two (one inner Cauchy and one external event) horizons. For $D>4$ the mass-horizon-radius relation, $M(r_{h},Q,D)$, exhibits always a minimum (unique for every value of the charge $Q$) which corresponds to the extreme BHs. This minimum arises at $r_{h}>0$ in all cases (no extreme black points). Besides these BH configurations there are also naked singularities, which arise when the mass $M$ lies below the minimum mass corresponding to the extreme BH solutions for a given charge.

The case of NEDs in $D=4$ dimensions, supporting elementary solutions which are bounded-strength electrostatic fields (for instance, the BI model) are exceptional. These models, which have been extensively analyzed in Ref.\cite{dr13}, support extreme and non-extreme black points and two kinds of single-horizon BHs (for large and small values of the charge), which are absent in the rest of families in $D=4$ and for all families in $D>4$ dimensions. For the asymptotically AdS cases, the same methods allow us to tackle the analysis of the BH structures, which are mainly determined by the short range behaviour of the solutions and rather unaffected by their asymptotic behaviours. We find BH configurations with similar qualitative properties, though their quantitative details  depend now also on the value of the cosmological constant length $l$.

Concerning the thermodynamic analysis of both asymptotically Schwarzschild and AdS black hole solutions (which only makes sense for non-IRD NEDs) we have first verified the fulfillment of the first law by generalizing the expressions of the mass and temperature. Subsequently we have explored the phase diagrams of the BH solutions associated to the different families in terms of the mass and charge of the configurations, as well as the behaviour of the temperature for these systems. The next step of our analysis was concerned with the existence of several relations among the different thermodynamic functions. Besides the finding of general Smarr formulae, valid for all admissible NED-supported asymptotically flat or AdS black holes, we have made use of the scale laws underlying the NED models. Indeed, the strategy of exploiting the scale symmetry of NEDs in flat spacetime has been extended to characterize the thermodynamic properties of the elementary BH solutions of G-NEDs in $D \geq 4$, both with asymptotically Schwarzschild or AdS behaviours. We have shown that scale symmetry is respected by the coupling to the gravitational field and we have found the representations of the one-parameter scale group in three-dimensional spaces built from trios of thermodynamic variables. The points of these spaces characterize the full set of BH states associated to the different models. In these spaces we have obtained the generating group equations (which are sets of linear, first-order, partial differential equations) as well as the associated trajectories. These trajectories form a ``universal" (NED-independent) beam of characteristic curves in every thermodynamic three-space. For a given G-NED the beam of characteristics generates a two-dimensional surface in this space. The points of the physically meaningful part of this surface are in a one-to-one correspondence with the full set of the associated elementary BH solutions (``equation of state"). This can be done starting from any non-characteristic curve whose points correspond to known solutions of the given model. In particular, by using the general equations obtained for extreme BH solutions in Sections \ref{sectionIII} and \ref{sectionIV}, we have derived the explicit general formulae giving the equation of state of the set of BH solutions, associated to any given (non-IRD) model, in the charge-entropy-temperature space, (once the explicit expression of the Lagrangian density function is specified). The knowledge of the equation of state allows to explore the thermodynamic structure (specific heats, phase transitions, etc.) of the set of elementary BH solutions associated to any admissible (but non-IRD) NED.

The bottom line of the research presented in this work is that the many results found in the literature for several NED models can be summarized and classified into a single framework, in such a way that a simple inspection of the vacuum and boundary behaviours of a given Lagrangian density $\varphi(X)$, following our methods, allows us to determine the main qualitative features of the associated elementary BH solutions, without any need of solving its particular field equations or using specific algebra for the analysis of the different metric and thermodynamic features. Moreover, the exact quantitative details can be obtained once the explicit expressions of the Lagrangian densities are replaced in the general formulae obtained, which finally allows the qualitative and quantitative analysis. These methods exhaust the set of admissible (non-IRD) NEDs, and severely constraint the new features that could be expected from future studies of such models in the context considered here.

It should be pointed out, however, that the methods and results presented here do not exhaust all the possibilities on this field. As examples, we mention the consideration of topological BHs, namely, BHs with flat or hyperbolic event horizons, or the asymptotically de-Sitter cases. Such scenarios would introduce further elements rendering the corresponding analysis much involved but, at the same time, opening the door to new and interesting issues. In addition, though our analysis is restricted to NEDs in General Relativity and many of the corresponding results would probably not survive to the extension of the gravitational field Lagrangian, or to the addition of non-minimal couplings between the matter and gravitational fields, it is nonetheless expected that similar methods as those developed here could still be applicable (see e.g. \cite{DRG15} for the case of Gauss-Bonnet gravity).

Several extensions of these methods to related problems deserve to be investigated. For instance, their generalization to the analysis of charged stationary axisymmetric BHs supported by NEDs would provide a more realistic description of the properties of astrophysical BHs in this context (see e.g. \cite{NEDregular8,Rodrigues:2017tfm,Lammerzahl:2018zvb} for some recent results at this regard). Also, the coupling to gravitation of other kinds of fields involving internal symmetries in flat space-time \cite{Volkov:1998cc} could benefit from the group techniques used here. Another interesting path would be to explore the applications of these methods to the framework of the AdS/CFT correspondence, within the consequences of the translation to the dual conformal field theory side of the symmetry properties of the G-NED models in AdS spacetime. This aspect has been largely studied in the literature for the case of the Born-Infeld electrodynamics \cite{BI-holo1,BI-holo2,BI-holo3,BI-holo4,BI-holo5,BI-holo6}. In this sense, the role to be played by the scale symmetry of the NED models and its associated scaling group equations in finding general new symmetries for conformal field theories is an open issue which deserves to be investigated. Work along several of the lines above is currently underway.

\section*{Acknowledgments} \label{Appendix}

DRG is funded by the Funda\c{c}\~ao para a Ci\^encia e a Tecnologia (FCT, Portugal) postdoctoral fellowship No.~SFRH/BPD/102958/2014, and acknowledges further support from the FCT research grants No. UID/FIS/04434/2013, No. PTDC/FIS-OUT/29048/2017 and No. PTDC/FIS-PAR/31938/2017, the Spanish projects  FIS2014-57387-C3-1-P (MINECO/FEDER, EU), FIS2017-84440-C2-1-P (AEI/FEDER, EU), the project H2020-MSCA-RISE-2017 Grant FunFiCO-777740, and the project SEJI/2017/042 (Generalitat Valenciana). We are indebted to Dr. Ph. Grandclement, Dr. R. B. Mann and Dr. J. A. Rodriguez-Mendez for useful comments.

\section*{Appendix} \label{Appendix}

We follow here the conventions of Ref. \cite{ortin} for the geometric analysis. The temporal and radial components of the Einstein tensor for the line element (\ref{eq:(3-5)}) are
\bea
{G_0}^{0}&=&\frac{1}{2\sqrt{\frac{\lambda\mu}{r^{D-2}}} } \frac{d}{dr} \left[\sqrt{\lambda \mu } (D-2)r^{D-3}\right]  \nonumber \\
&+& \frac{D-2}{2r} \sqrt{\lambda \mu} \frac{d}{ dr} \left( \sqrt{\frac{\mu }{\lambda}} \right) - \frac {(D-2)(D-3)}{2r^{2}}
\label{eq:(A-1)} \\
{G_1}^{1} &=& -\frac{D-2}{2r} \sqrt{\lambda\mu} \frac{d}{dr} \left( \sqrt{\frac{\mu}{\lambda}} \right) - \frac{(D-2)(D-3)}{2r^{2}}
\nonumber \\
&+&
\frac{1}{2\sqrt{\frac{\lambda  }{\mu}} {r^{D-2}}} \frac{d}{dr} \left[\sqrt{\lambda\mu} (D-2)r^{D-3}\right] \ ,
\label{eq:(A-2)}
\ena
whereas the remaining components are
\bea
{G_0}^{i} &=& {G_i}^{0} = 0 \\
\label{eq:(A-3)}
{G_i}^{j} &=& \delta_{i}^{j}  \Big\lbrace \Big(  \frac{-1}{r^{D-2}\sqrt{\frac{\lambda }{\mu}} }
\frac{ d }{ dr } \Big[ r^{D-3} \sqrt{ \lambda \mu  }\Big( 1-\frac{ D-2 }{ 2 } \Big) \Big] \nonumber \\
&+&
\frac{ D-3 }{r^{2}} \Big( 1-\frac{D-2}{2} \Big) \Big)+ \frac{ D-2 }{ 2r } \sqrt{ \lambda \mu  } \frac{ d }{ dr } \Big( \sqrt{ \frac{ \mu  }{ \lambda  }  } \Big) \nonumber \\
&+&
\frac{ 1 }{ 2\sqrt{ \frac{ \lambda  }{ \mu  }  } r^{ D-2 } } \frac{ d }{ dr } \Big( r^{ D-2 }\sqrt{ \frac{\mu}{\lambda} } \frac{ d\lambda  }{ dr } \Big) \Big\rbrace \ .
\label{eq:(A-4)}
\ena

We can rewrite the independent mixed components of the Einstein tensor in the Schwarzschild coordinates of Eq.(\ref{eq:(3-14)}) as
\bea
{ G_0 }^{ 0 } &=& {G_1}^{1} = \frac { 1 }{ 2{ r }^{ D-2 } } \frac { d }{ dr } \left( g(r) (D-2){ r }^{ D-3 }\right) \label{eq:(A-5)} \\
&-&
\frac { (D-2)(D-3) }{ 2{ r }^{ 2 } } \nonumber \\
{G_p }^{ q } &=& \delta_{ p }^{ q } \Big\lbrace \frac { -1 }{ { r }^{ D-2 } } \frac { d }{ dr }
\Big[ \left(2-\frac { D }{ 2 } \right) g(r) { r }^{ D-3 }\Big] \label{eq:(A-6)} \\
&+& \frac { D-3 }{ { r }^{ 2 } } \Big( 2-\frac { D }{ 2 } \Big) + \frac { 1 }{ 2{ r }^{ D-2 } } \frac { d }{ dr }
\Big( { r }^{ D-2 }\frac { d g(r)  }{ dr } \Big) \Big\rbrace \ . \nonumber
\ena
From these expressions the Einstein equations with cosmological term become
\bea
&\frac{d}{dr}& \left[(g(r) - 1)r^{D-3}\right] + \Lambda r^{D-2} = -\frac{2\chi}{D-2} r^{D-2} {T_0}^{0} \nonumber \\
&=& -\frac{2\chi}{D-2} \left( 2 Q E(r) - r^{D-2} \varphi \right) \ ,
\label{eq:(A-7)}
\ena
and
\bea
&\frac{d}{dr}& \Big[r^{D-2}\frac{dg(r)}{dr} + (D-4) g(r) r^{D-3}\Big] + (D-2)\Lambda r^{D-2}\nonumber \\
&=& -2\chi r^{D-2} {T_2}^{2} = 2\chi r^{D-2} \varphi \ .
\label{eq:(A-8)}
\ena

The compatibility of these equations can be straightforwardly established. Let us thus work with the first one (\ref{eq:(A-7)}). This equation can be formally integrated in the general case. Indeed, if we integrate both sides between two radii $r_{1}$ and $r_{2}$, we obtain
\bea
&&\Big[(g(r) - 1)r^{(D-3)}\Big]\Big\vert_{r_{1}}^{r_{2}} + \frac{\Lambda r^{(D-1)}}{{D-1}}\Big\vert_{r_{1}}^{r_{2}}  \label{eq:(A-9)}  \\
&-&\frac{2\chi}{D-2} \int_{r_{1}}^{r_{2}} dr r^{D-2} {T_0}^{0}
= -\frac{2\chi \cdot \varepsilon(r_{1},r_{2},Q,D) }{(D-2)\omega_{(D-2)}} \ , \nonumber
\ena
where $\varepsilon(r_{1},r_{2},Q,D)$ is the field energy contained in the space between the hyper spheres $S^{D-2}(r_{1})$ and $S^{D-2}(r_{2})$ (see Eq.(\ref{eq:(2-30)})). From this last equation, setting $r_2 \rightarrow \infty$ and $r_1=r$, the expression (\ref{eq:(3-20)}) is immediately found for asymptotically AdS solutions.

\end{document}